\documentclass[a4paper,11pt]{article}
\pdfoutput=1 

\usepackage{jheppub} 
            
\usepackage{bm,amssymb,slashed,graphicx,multirow,soul,mathtools,xspace,array}  
\usepackage{float}                   
\allowdisplaybreaks
\usepackage{ bbold }
\usepackage{subfigure}

\newcounter{RSQ}

\newcounter{Duff}

\newcommand{\todo}[1]{{\color{red} \ifmmode\else[todo]\fi #1}}
\usepackage[usenames,dvipsnames]{xcolor}
     \definecolor{hgreen}{rgb}{0,.3,0}
     \definecolor{darkgreen}{rgb}{0.3,.8,0.2}
     \definecolor{hred}{rgb}{.3,0,0}
     \definecolor{hblue}{rgb}{0,0,.3}
     \definecolor{LightGray}{gray}{0.95}
     
\usepackage{hyperref}

\newcommand{\zh}{z_h} 
\newcommand{\ph}{p_h} 
\renewcommand{\figurename}{\bf Figure}
\renewcommand{\tablename}{\bf Table}
\def\bea#1\eea{\begin{align}#1\end{align}}
\newcommand{\nn}{\nonumber\\}
\newcommand{\bef}{\begin{figure}[h!tb]\centering}
\newcommand{\eef}{\end{figure}}

\usepackage{mathrsfs}

 \usepackage{etoolbox}
    \makeatletter
    \patchcmd{\maketitle}{\@fpheader}{}{}{}
    \makeatother

\newcommand\snowmass{
\begin{center}
  \rule[-0.2in]{\hsize}{0.01in}\\
  \rule{\hsize}{0.01in}\\
  \vskip 0.1in
  Submitted to the Proceedings of the US Community Study\\ 
  on the Future of Particle Physics (Snowmass 2021)\\
  \rule{\hsize}{0.01in}\\
  \rule[+0.2in]{\hsize}{0.01in}\\[-2em]
\end{center}
}

\def\beq{\begin{equation}}
\def\eeq{\end{equation}}

\title{Snowmass 2021 White Paper: Resummation for future colliders}

\author[a]{Melissa van Beekveld,}
\author[b]{Sebastian Jaskiewicz,}
\author[c,d]{Tao Liu,}
\author[e,f]{Xiaohui Liu,}
\author[g]{Duff Neill,}
\author[h]{Alexander Penin,}
\author[i,j]{Felix Ringer,}
\author[k]{Robert Szafron,}
\author[l,m]{Leonardo Vernazza,}
\author[n]{Gherardo Vita,}
\author[o]{Jian Wang}

\affiliation[a]{Rudolf Peierls Centre for Theoretical Physics, Clarendon Laboratory, Parks Road, University of Oxford, Oxford OX1 3PU, UK}
\affiliation[b]{Institute for Particle Physics Phenomenology, Durham University, Durham DH1 3LE, United Kingdom}
\affiliation[c]{Institute of High Energy Physics,  Chinese Academy of Sciences, Beijing 100049, China}
\affiliation[d]{University of Chinese Academy of Sciences, Beijing 100049, China}
\affiliation[e]{Center of Advanced Quantum Studies, Department of Physics,
Beijing Normal University, Beijing 100875, China} 
\affiliation[f]{Center for High Energy Physics, Peking University, Beijing 100871, China}
\affiliation[g]{Theoretical Division, MS B283, Los Alamos National Laboratory, Los Alamos, NM 87545, USA}
\affiliation[h]{Department of Physics, University of Alberta, Edmonton AB T6G 2J1, Canada}
\affiliation[i]{C.N. Yang Institute for Theoretical Physics, Stony Brook University, Stony Brook, NY 11794,
USA}
\affiliation[j]{Department of Physics and Astronomy, Stony Brook University, Stony Brook, NY 11794, USA}
\affiliation[k]{Department of Physics, Brookhaven National Laboratory, Upton, N.Y., 11973, USA}
\affiliation[l]{INFN, Sezione di Torino, Via P. Giuria 1, I-10125 Torino, Italy}
\affiliation[m]{Nikhef, Science Park 105, NL-1098 XG Amsterdam, The Netherlands}
\affiliation[n]{SLAC National Accelerator Laboratory, Stanford University, Stanford, CA 94039, USA}
\affiliation[o]{School of Physics, Shandong University, Jinan, Shandong 250100, China}

\emailAdd{melissa.vanbeekveld@physics.ox.ac.uk}
\emailAdd{sebastian.jaskiewicz@durham.ac.uk}
\emailAdd{liutao86@ihep.ac.cn}
\emailAdd{xiliu@bnu.edu.cn}
\emailAdd{duff.neill@gmail.com}
\emailAdd{penin@ualberta.ca}
\emailAdd{felix.ringer@stonybrook.edu}
\emailAdd{rszafron@bnl.gov}
\emailAdd{leonardo.vernazza@to.infn.it}
\emailAdd{gvita@stanford.edu}
\emailAdd{j.wang@sdu.edu.cn}

\date{\today}

\abstract{Resummation techniques are essential for high-precision phenomenology at current and future high-energy collider experiments. Perturbative computations of cross sections often suffer from large logarithmic corrections, which must be resummed to all orders to restore the reliability of predictions from first principles. The precise understanding of the all-order structure of field theories allows for fundamental tests of the Standard Model and new physics searches. In this white paper, we review recent progress in modern resummation techniques and outline future directions. In particular, we focus on the resummation beyond leading power, the joint resummation of different classes of logarithms relevant for jets and their substructure, small-$x$ resummation in the high-energy regime and  the QCD fragmentation process in the small-$z_h$ limit. }

\begin{document}
\preprint{
 IPPP/22/11, 
 SLAC-PUB-17664,
 YITP-SB-2022-06
}

\snowmass

 \maketitle
 
\vfill

\pagebreak

\section{Introduction}
Tests of the Standard Model and New Physics searches at the Large Hadron Collider (LHC), Electron-Ion Collider (EIC) and other future colliders require extremely precise theoretical predictions. Correctly accounting for large quantum corrections will be indispensable for the High-Luminosity phase of the LHC, especially since the direct searches for new particles have not resulted in a discovery so far. Precise measurements of the Higgs boson and top quark properties, or tests of the Standard Model's flavor structure, rely on an excellent theoretical control over the QCD-induced corrections.  

The accurate theoretical description of high-energy processes at colliders is among the most successful applications of perturbative QCD. The relevant QCD corrections are typically computed as an expansion in the strong coupling constant $\alpha_s$. Currently, state-of-the-art predictions include corrections up to the third order in $\alpha_s$, with most of the processes being known only up to the next-to-next-to-leading order (NNLO). 

A typical process in QCD depends on various energy scales. When the scales are widely separated, the presence of the scale hierarchy manifests itself through logarithmically enhanced radiative corrections, which require an all-order resummation of the series in the strong coupling constant. One of the great simplifications of perturbative analysis  at high energy is that the contributions suppressed by a power of a ratio of an infrared scale of the process to its center-of-mass energy are small and can be neglected in the first approximation, reducing the number of scales and hence the complexity of the calculations. Thus, the resummation techniques typically target the leading power (LP) terms in the power expansion. Before we review in more detail the recent progress in the most challenging fields, such as subleading power resummation, small-$x$ resummation, fragmentation and small-$z_h$ resummation, and the physics of jets and their substructure, let us briefly mention two of the well-established application of resummation techniques, which played a crucial role in the development of this research field.  

Threshold resummation is one of the flagship applications of factorization and resummation for collider physics \cite{Parisi:1979xd,Curci:1979am,Sterman:1986aj,Catani:1989ne,Catani:1990rp}. Typically, when the phase space available for radiation is limited,  the corrections are dominated by soft gluons. Therefore, threshold resummation is relevant for the production of heavy states. In the context of LHC physics, the leading power threshold resummation is well understood. It has been applied to many processes, such as the production of the electroweak gauge and Higgs bosons and top quarks as well as production of new heavy particles, see e.g. Refs.~\cite{Moch:2005ky,Laenen:2005uz,Idilbi:2005ni,Becher:2007ty,Kulesza:2008jb,Beneke:2011mq,Bonvini:2012az,Bonvini:2014joa,Hinderer:2018nkb}. 

A second important class of problems involves transverse momentum resummation. Substantial progress has been achieved in recent years, with the state-of-the-art results reaching next-to-next-to-next-to leading logarithmic (N$^3$LL) accuracy combined with NNLO and even N$^3$LO fixed order computations, see e.g. \cite{Li:2016ctv,Bizon:2017rah,Coradeschi:2017zzw,Chen:2018pzu,Camarda:2019zyx,Bacchetta:2019sam,Luo:2019szz,Ebert:2020yqt,Kallweit:2020gva,Wiesemann:2020gbm,Ebert:2020dfc,Luo:2020epw,Becher:2020ugp,Ebert:2020qef,Ebert:2020sfi,Ju:2021lah,Neumann:2021zkb,Billis:2021ecs,Chen:2022cgv}. These results are very important components for precision tests of the Standard Model at the LHC. 

The need for a precise understanding of the all-order structure of QCD led to the development of new theoretical techniques and methods. Today the field of resummation is exploring new topics crucial for phenomenology and our understanding of gauge theories. This paper reviews the recent progress in the theoretical methods of resummation and their relevance for collider phenomenology. We will discuss the most challenging problems currently under investigation, which are addressed with soft-collinear effective field theory (SCET) \cite{Bauer:2000yr,Bauer:2001yt,Beneke:2002ph,Beneke:2002ni} and directly in QCD. In section \ref{sec:NLP}, we discuss recent progress in understanding the all-order structure of subleading power terms using the traditional diagrammatic approach and SCET. Section \ref{sec:jets} provides an overview of jet physics and jet substructure observables. Here the focus is in particular on the simultaneous resummation of several classes of large logarithmic corrections. This aspect is particularly relevant for jet substructure observables, where multiple energy scales appear. Next, in section \ref{sec:smallx}, we discuss recent progress in understanding gluon saturation, which is relevant in the high-energy regime where the initial-state gluon momentum fraction is small. Section \ref{sec:fragmentation} focuses on the QCD fragmentation process in the limit where the energy fraction of the identified hadrons becomes small relative to the hard scale of the process. Finally, we conclude in section \ref{sec:conc}.

\section{Resummation beyond leading power}
\label{sec:NLP}

In the leading-power approximation the
structure of the logarithmic corrections is in general well
understood and the relevant resummation techniques are well
elaborated. However, in many crucial cases the continually
improving accuracy of the experimental measurements requires
the inclusion of power suppressed terms in the
theoretical estimates. As a result, much effort is currently
invested into the study of a diverse class of power
corrections, see \cite{Penin:2014msa,Melnikov:2016emg,Penin:2016wiw,Liu:2017axv,Liu:2017vkm,Boughezal:2018mvf,Bruser:2018jnc,Moult:2018jjd,Ebert:2018lzn,Liu:2018czl,Beneke:2018gvs,Engel:2018fsb,Ebert:2018gsn,Bahjat-Abbas:2019fqa,Penin:2019xql,Beneke:2019mua,Beneke:2019oqx,Anastasiou:2020vkr,Liu:2020wbn,Broggio:2021fnr,Liu:2021chn}
and references therein. Incorporating logarithmically
enhanced power-suppressed terms can significantly increase the
accuracy and extend the region where the perturbative calculation is applicable. Besides the phenomenological
importance, the  power-suppressed contributions are very
interesting from the general effective field theory point of
view since the structure of the factorization and
renormalization group evolution in this case becomes highly
nontrivial already in the leading logarithmic (LL) approximation.

\subsection{Diagrammatic methods for threshold resummation}

\paragraph{Large logarithms near threshold.}
Power corrections play an important role for precise 
predictions of physical observables involving heavy final 
states. Because of a phenomenon of dynamical enhancement, 
these scattering processes receive a large contribution 
near their partonic threshold, where the dimensionless 
partonic cross takes the following schematic form 
\begin{equation}
\label{threshold_simple}
    \Delta(\xi) = \sum_{n=0}^{\infty}\alpha_s^n 
    \Bigg\{ \sum_{m=0}^{2n-1} c^{\rm LP}_{nm} 
    \left(\frac{\ln^m \xi}{\xi}\right)_+ + d_n\delta(\xi)
    + \sum_{m=0}^{2n-1} c^{\rm NLP}_{nm} \ln^m \xi 
    + \dots \Bigg\}\,.
\end{equation}
In this equation $\xi$ represents a variable, which 
approaches $\xi \to 0$ near threshold; for instance, 
$\xi = 1-z \equiv 1-Q^2 /\hat s$ for the Drell-Yan process, where 
$Q^2$ is the invariant mass of the final state, and 
$\hat s$ the partonic center of mass energy squared. For 
$\xi \to 0$ the partonic cross section in 
Eq.~({\ref{threshold_simple}}) develops large 
logarithms that need to be resummed to obtain 
precise predictions. The terms proportional to 
$c^{\rm LP}_{nm}$ contribute at leading power, 
and their resummation has been known for 
a long time to a high logarithmic accuracy, since the seminal papers~\cite{Parisi:1979xd,Curci:1979am,Sterman:1986aj,Catani:1989ne,Catani:1990rp}.
Later, LP threshold resummation has been reinterpreted 
and clarified using a wide variety of methods, including 
the use of Wilson lines~\cite{Korchemsky:1992xv,Korchemsky:1993uz}, 
the renormalization group~\cite{Forte:2002ni}, the 
connection to factorization theorems~\cite{Contopanagos:1996nh}, 
and soft collinear effective theory~\cite{Becher:2006nr,Schwartz:2007ib,Bauer:2008dt}. 
The state-of-the-art for resummation at LP is 
next-to-next-to-next-to-leading logarithmic (N$^3$LL)
accuracy for color singlet final states, and 
next-to-next-to-leading logarithmic (NNLL)
accuracy for processes involving colored 
particles in the final state.

The resummation of logarithms with $c^{\rm NLP}_{nm}$ coefficients, 
contributing at next-to-leading power (NLP), has been 
considered only more recently and it is still subject 
of intense study. Progress has been made by employing 
diagrammatic~\cite{Bahjat-Abbas:2019fqa,vanBeekveld:2019prq,vanBeekveld:2021hhv,vanBeekveld:2021mxn}, 
effective field theory~\cite{Moult:2018jjd,Beneke:2018gvs,Beneke:2019mua,Moult:2019uhz,Beneke:2020ibj} 
and renormalization group methods~\cite{Ajjath:2020ulr,Ajjath:2020sjk}. 
In this section, we provide a brief overview of the 
diagrammatic approach, while the application of SCET 
to the resummation of large logarithms at NLP will 
be discussed in the following section. 

\paragraph{Resummation at NLP for diagonal partonic channels.}
Near threshold, additional radiation is constrained 
to be soft and the partonic cross section develops 
a hierarchy of scales, with $Q$ representing the 
energy involved in the hard interaction, and $Q(1-z)$ 
with $z \ll 1$ the energy of the soft radiation.
It can be shown that soft radiation at LP 
is described in terms of uncorrelated eikonal 
emissions. As a consequence, the partonic cross 
section factorizes into a soft and a hard function, 
the latter representing the central hard scattering, 
while the former describes soft radiation 
in terms of a vacuum expectation value
of Wilson lines along the directions of the 
hard scattering particles. The exponentiation of 
eikonal radiation together with the factorization
of the associated phase space (in Mellin space) 
is the basis for 
resummation of the logarithms proportional 
to the $c^{\rm LP}_{nm}$ coefficients  in 
Eq.~(\ref{threshold_simple}).

Beyond LP, the factorization of the partonic 
cross section becomes more involved. In general, 
this happens because soft radiation becomes 
sensitive to the nature of the hard scattering 
particles and their hard interaction. For 
instance, soft gluons can be emitted through 
chromo-magnetic interactions, which are 
sensitive to the spin of the emitting particles; 
furthermore, at NLP one needs to take into 
account the emission of soft quarks, too. 
In this section we will focus on the emission 
of soft gluon radiation from collinear legs, which contributes at 
LP and NLP to the large logarithms in diagonal 
production channels, such as $q\bar q$ in 
Drell-Yan, or $gg$ in Higgs production. 
The emission of soft quarks from collinear legs
contributes instead to the so-called off-diagonal 
partonic channels, such as the $qg$ channel 
in Drell-Yan. We will briefly discuss these 
channels in the next subsection. 

Besides being sensitive to the spin of the 
emitting particles, soft gluon radiation 
at NLP begins to reveal the structure of the 
hard scattering~\cite{Low:1958sn,Burnett:1967km}, 
as well as the structure of virtual collinear 
radiation associated to the hard scattering 
particles of the process~\cite{DelDuca:1990gz}. 
These effects can be described in terms of 
convolutions of soft and collinear matrix 
elements multiplying a hard function, 
which represents the central non-radiative 
scattering process. In a diagrammatic approach, 
the soft and collinear matrix elements are 
described in terms of QCD fields and Wilson 
lines. For instance, in the case of Drell-Yan, 
the $q\bar q$ amplitude with an emission of 
one additional soft gluon is written 
(up to NNLO) as~\cite{Bonocore:2015esa,Bonocore:2016awd}
\begin{equation} \label{NEfactor_nonabel2}
{\cal A}_{\mu, a} (p_j, k)  =  
\sum_{i = 1}^2 \left\{ \frac12 \, 
\widetilde{\cal S}_{\mu, a} (p_j, k) 
+ g \, {\bf T}_{i, a} \, G^\nu_{i, \mu} \, 
\frac{\partial}{\partial p_i^\nu} 
+ \big[ J_{\mu, a} \left(p_i, n_i, k\right)
- {\cal J}_{\mu, a} \left(p_i, n_i, k\right) \big] \right\}
{\cal A} (p_j) \, ,
\end{equation} 
where $p_j$, $j = 1,2$ represent the momenta
of the initial state partons, and 
\begin{equation} 
G^{\mu \nu}_i = g^{\mu \nu} - 
\frac{(2 p_i - k)^\nu}{2 p_i \cdot k - k^2} \, k^\mu.
\end{equation} 
In Eq.~(\ref{NEfactor_nonabel2}), 
$\widetilde{\cal S}_{\mu, a} (p_j, k)$ represents 
a generalized soft function, which accounts for 
soft emissions from the external particles beyond 
the eikonal approximation~\cite{Laenen:2008gt,Laenen:2010uz}; 
the second term is given by the orbital angular 
momentum operator acting on the non-radiative 
amplitude ${\cal A} (p_j)$, and represents the 
so-called Low-Burnett-Kroll (LBK) theorem \cite{Low:1958sn,Burnett:1967km}. In the last 
term, $J_{\mu, a} \left(p_i, n_i, k\right)$ 
represents a radiative jet function, introduced 
first in \cite{DelDuca:1990gz}, while 
${\cal J}_{\mu, a} \left(p_i, n_i, k\right)$ 
subtracts overlapping soft modes between the 
radiative and the soft function.

Starting from Eq.~(\ref{NEfactor_nonabel2}) it is possible
to derive the corresponding factorized cross section, 
which has been explicitly verified up to NNLO in 
perturbation theory~\cite{Bonocore:2015esa,Bonocore:2016awd}.
The factorization theorem in Eq.~(\ref{NEfactor_nonabel2})
is valid up to this perturbative order, which means that 
it is not enough to develop a theory of resummation at NLP, 
at arbitrary logarithmic accuracy. While studies of the all-order
factorization structure are in development (see e.g.~\cite{Gervais:2017yxv,Laenen:2020nrt,Bonocore:2021cbv}
for recent progress), it has already been  possible to 
exploit the result of Eq.~(\ref{NEfactor_nonabel2}) 
to obtain the resummation of NLP large logarithms 
at leading logarithmic accuracy. Indeed, it is
possible to show that the second and third terms in 
Eq.~(\ref{NEfactor_nonabel2}) contribute only starting 
at next-to-leading logarithmic (NLL) accuracy, i.e., 
leading logarithms are generated only by momenta 
configurations which are maximally soft and collinear 
(see~\cite{Bahjat-Abbas:2019fqa} for a detailed 
discussion), which are taken into account by the 
generalized soft function, i.e. the first term in
Eq.~(\ref{NEfactor_nonabel2}). As a consequence, 
at LL accuracy the factorization theorem simplifies. 
In Mellin space, the bare partonic cross 
section takes the form 
\begin{equation}
  \widehat{\Delta}^{(q \bar{q})}_{\rm NLP,LL} \left( N, Q^2, \epsilon \right) \, = \, 
  \left| \mathcal{H} \left( Q^2 \right) \right|^2
  {\cal S} \! \left( N, Q^2, \epsilon \right) \, .
\label{sigmahadN2}    
\end{equation}
By using tools such as the replica trick \cite{Gardi:2010rn},
the generalized soft function ${\cal S} \! \left( N, Q^2, 
\epsilon \right)$ can be shown to exponentiate,~\cite{Laenen:2010uz},
leading to~\cite{Bahjat-Abbas:2019fqa}
\begin{equation}
\label{NEresum}
\int_0^1 dz z^{N - 1} \Delta_{\rm \, NLP,LL}^{(q \bar{q})}(z) \, 
= \, {\Delta}^{(q \bar{q})}_{\rm LO} (Q^2) 
\exp \left[ \frac{2 \alpha_s C_F}{\pi} 
\left(\log^2 \! N + \frac{\log N}{N} \right) \right] \, ,
\end{equation}
where the exponent of the one-loop soft function resums 
large leading logarithms, with the second term corresponding 
to the NLP correction. 

It can be proven that the structure of the NLP correction in Eq.~(\ref{NEresum}) 
is universal for all electroweak annihilation processes involving a colorless final state \cite{DelDuca:2017twk,Bahjat-Abbas:2019fqa}. The LL resummed cross 
section, up to NLP, takes the form of Eq.~(\ref{NEresum}),
with the LO partonic cross section multiplying a universal 
factor, whose color structure ($C_F$ in Eq.~(\ref{NEresum})) 
depends on the initial state particles. In general, in $z$ 
space this leads to the resummation formula 
\begin{eqnarray} \label{eq:resumgeneral} \nonumber
\Delta_{aa}^{\rm dQCD,\,LP+NLP}(N,Q^2/\mu^2) 
&=& g_{0}(\alpha_s)\,{\rm exp}\Bigg\{\int_0^1{\rm d}z\,z^{N-1}\,\Bigg[ \frac{1}{1-z}D_{aa}\left(\alpha_s\left(\frac{(1-z)^2Q^2}{z}\right)\right) \\
&&\hspace{1.0cm} + 2 \int_{\mu^2}^{(1-z)^2Q^2/z}\frac{{\rm d}k_T^2}{k_T^2} P^{\rm LP+NLP}_{aa}\left(z,\alpha_s(k_T^2)\right)\Bigg]_+\,\Bigg\},
\end{eqnarray}
where $g_{0}(\alpha_s)$ collects the $N$-independent 
contributions. Large logarithms are resummed in the exponent, 
which contains the diagonal DGLAP splitting function 
$P_{aa}(z,\alpha_s)$ and soft wide-angle contributions, 
collected in $D_{aa}(\alpha_s)$, where $aa = q\bar q$ 
for quark-antiquark initiated processes, and $aa = gg$ 
for gluon-gluon initiated processes. Evaluating these 
functions to a given order in $\alpha_s$ resums the large 
logarithms to a corresponding logarithmic accuracy. To 
date, $P_{aa}(z,\alpha_s)$ is known to fourth order in 
$\alpha_s$ and $D_{aa}(\alpha_s)$ to third order, which 
guarantees the resummation (at LP) up to N$^3$LL accuracy. 
Both terms are process-independent, to the extent that 
they only depend on the color structure of the underlying 
hard-scattering process. The NLP correction in
Eq.~(\ref{NEresum}) is given by expanding 
$P_{aa}(z,\alpha_s)$ up to NLP in the threshold variable. 
This ensures that all LLs are resummed at NLP. In principle, Eq.~(\ref{eq:resumgeneral}) resums also NLP logarithms 
beyond LL accuracy. These originates from keeping higher 
order contributions in $P_{aa}(z,\alpha_s)$ at NLP, as 
well as from the argument of $\alpha_s$ in the function 
$D_{aa}$ and from the $1/z$-dependence of the upper limit 
of the $k_T$ integral. However, these contributions give 
only part of the whole NLL contribution and beyond, 
and can be dropped for a consistent resummation at 
fixed logarithmic accuracy. 

\begin{figure}[t]
\begin{center}
\includegraphics[width=0.48\textwidth]{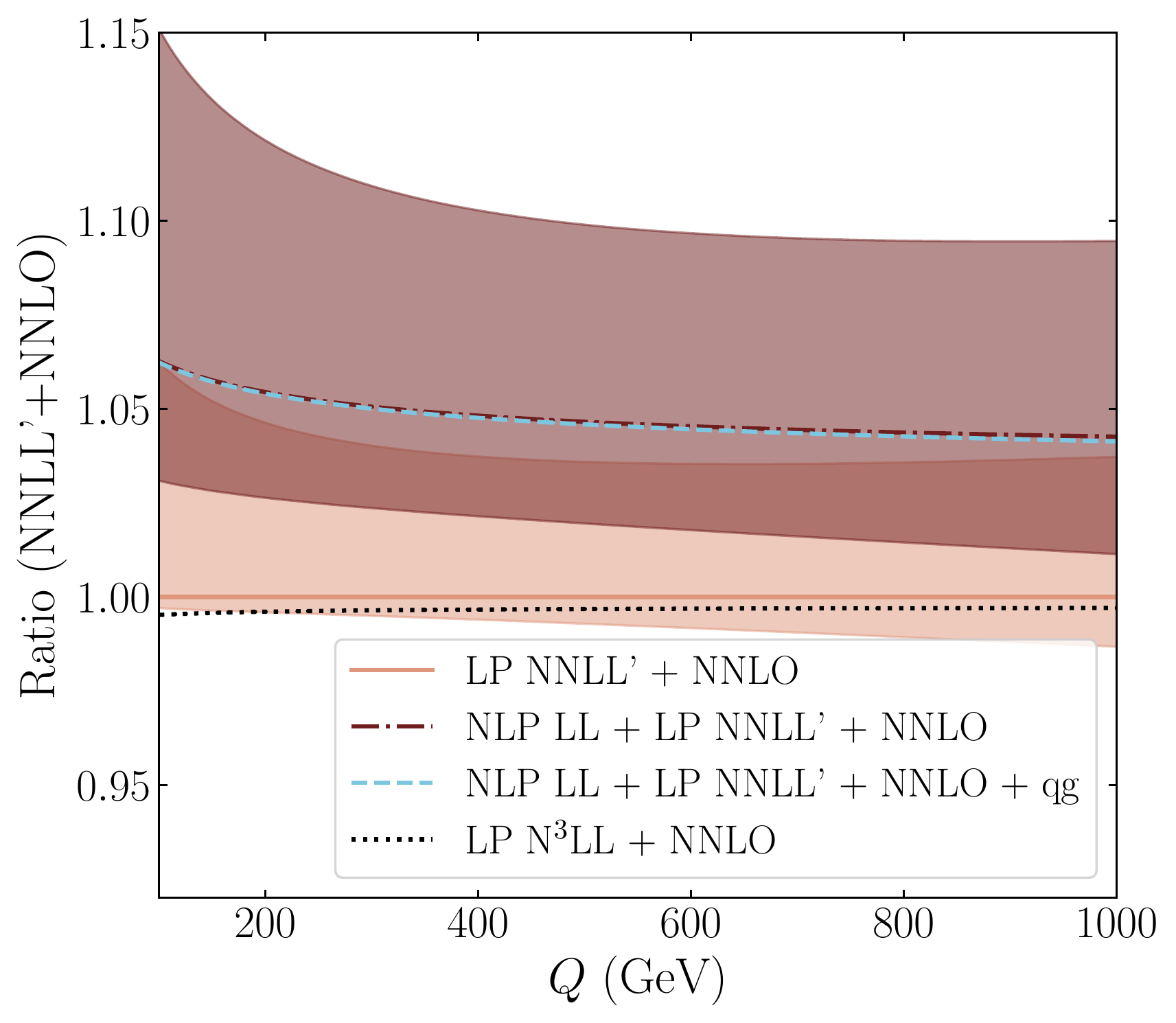}
\includegraphics[width=0.48\textwidth]{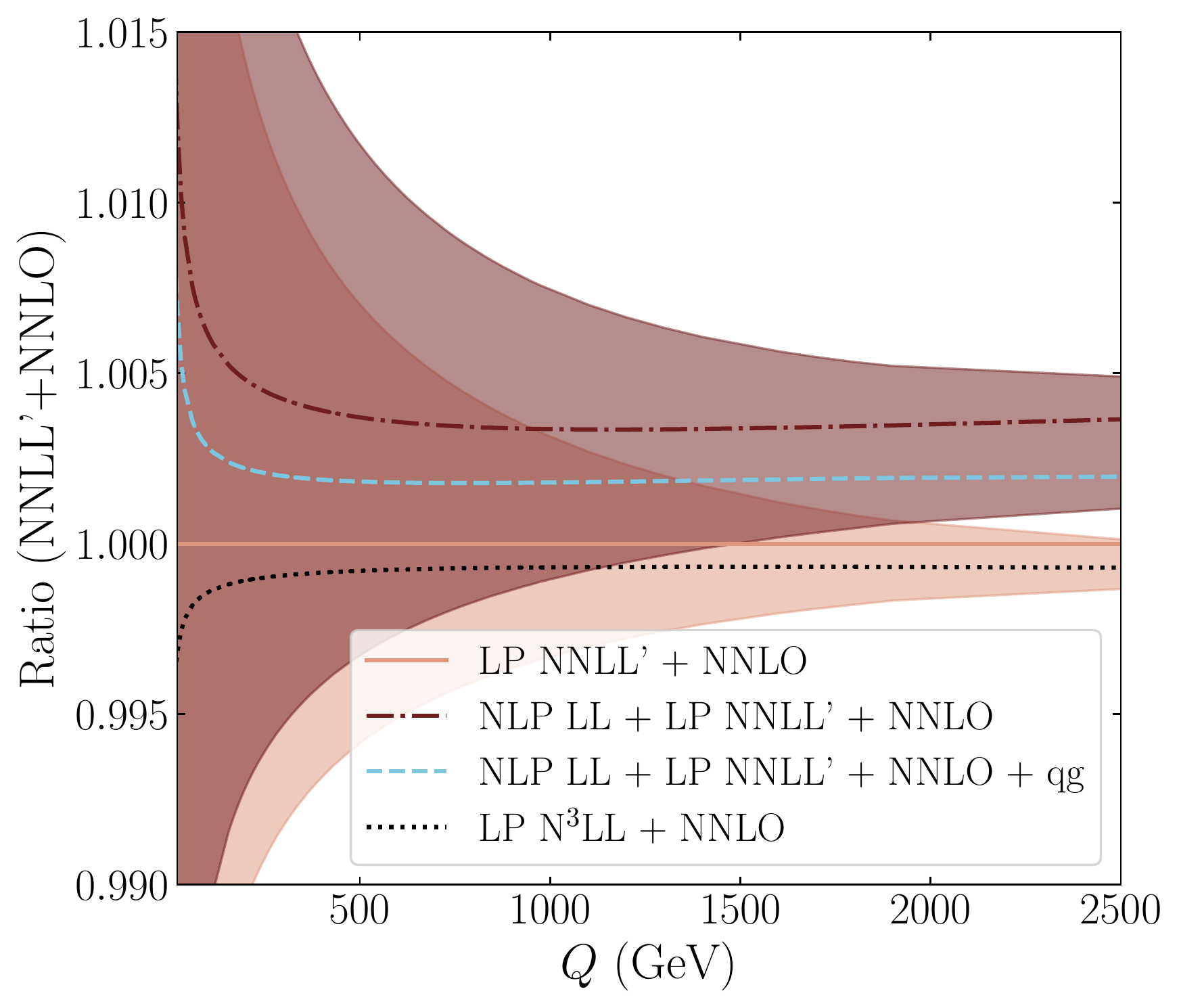}
\end{center}
\caption{Ratio plot of the total Higgs cross section 
(left) and the DY invariant mass distribution (right), 
normalized to the  NNLL$^{\prime}$ + NNLO result. 
The NNLL$^{\prime}$ (+NLP LL) result is again shown 
by the orange solid (red dash-dotted) lines. The 
NNLL$^{\prime}$ NLP LL result is obtained by adding 
the N$^3$LO NLP LL $qg$ result, which is shown by the 
dotted light-blue line. Plots taken from \cite{vanBeekveld:2021hhv}.}\label{fig:NNLLratios}
\end{figure} 
An extensive phenomenological analysis investigating the 
effects of resumming threshold LLs at NLP in physical 
observables of interest for the LHC has been presented 
in~\cite{vanBeekveld:2021hhv}. There, it has been
shown that, in general, the NLP LL correction becomes 
competitive with the resummation of LP logarithm
at NNLL accuracy, as can be seen for instance in 
Fig.~\ref{fig:NNLLratios}. Therefore, any precise 
prediction in which LP resummation is performed at 
NNLL accuracy and beyond should include 
resummation of NLP logarithms. The analysis of~\cite{vanBeekveld:2021hhv} has also shown 
that the diagrammatic and the effective field 
theory (SCET) approach~\cite{Beneke:2018gvs,Beneke:2019mua}, 
give almost identical results, when the same 
contributions at NLP are taken into account, 
thus removing earlier claims of numerical 
differences between the two approaches. 

\paragraph{Resummation at NLP for 
subleading partonic channels.}
\begin{figure}[t]
\begin{center}
\includegraphics[width=0.3\textwidth]{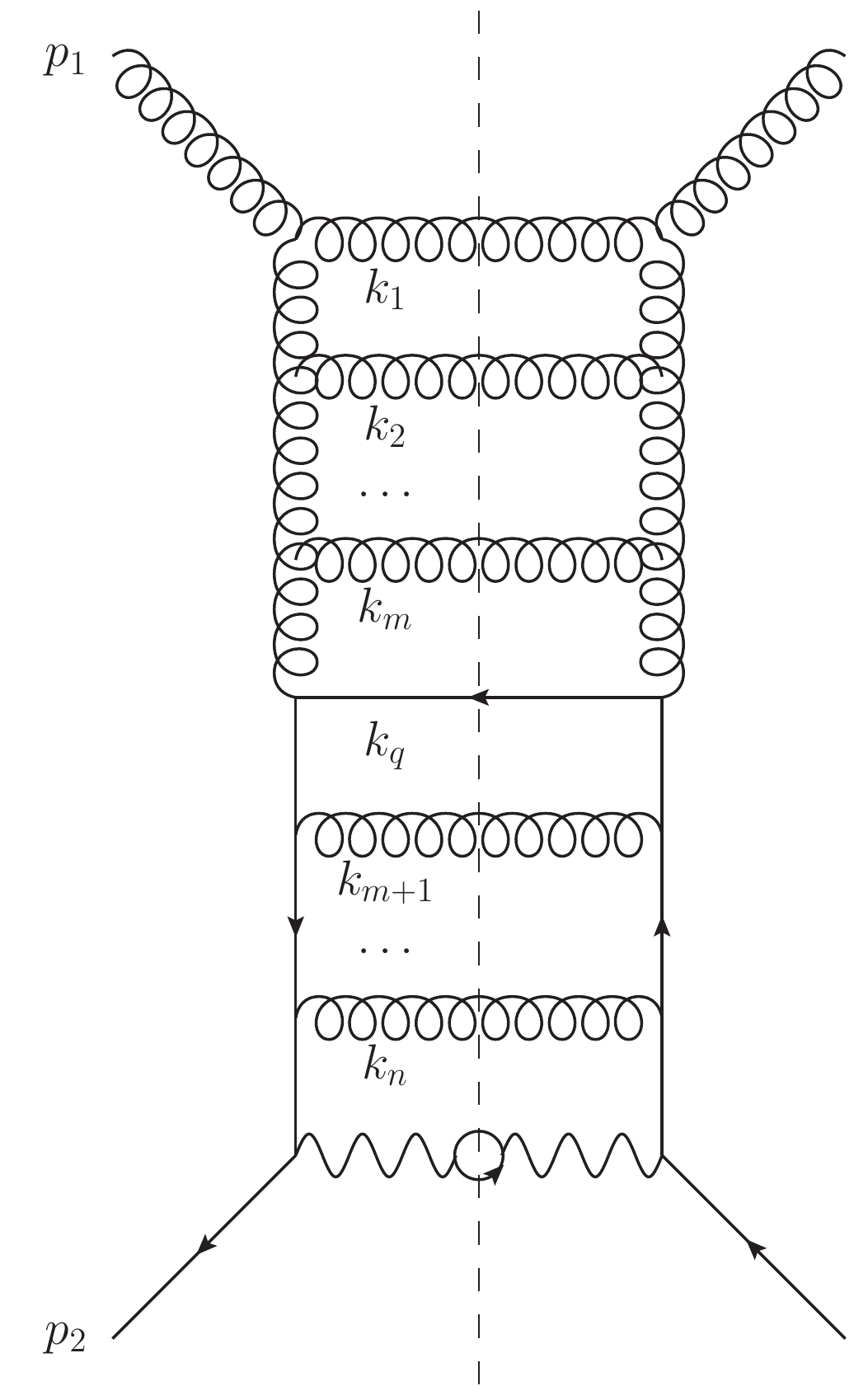}
\qquad \qquad \qquad
\includegraphics[width=0.3\textwidth]{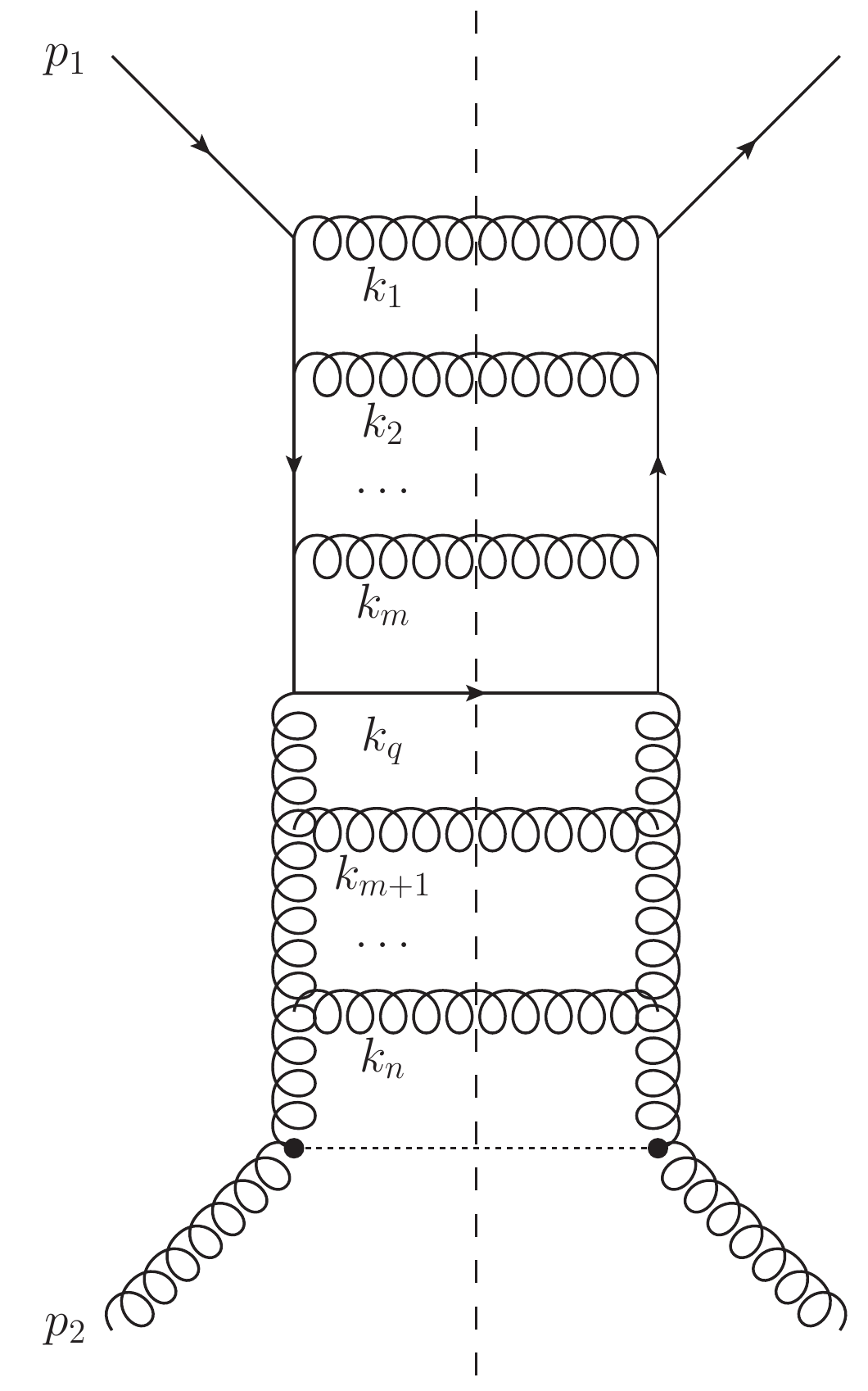}
\caption{Left: ladder diagram contributing to the $g\bar{q}$ 
channel in DY production; right: similar but for the $qg$ 
channel in Higgs production.}
\label{fig:DYladder}
\end{center}
\end{figure}
The resummation of large threshold logarithms 
in off-diagonal channels, such as
the $qg$ channel in Drell-Yan, are particularly 
interesting. These channels start contributing already at 
NLP, because near threshold they involve the emission
of a soft quark, which is itself power-suppressed with respect to a soft gluon 
emission. Earlier conjectures for Deep Inelastic 
Scattering (DIS) and Drell-Yan \cite{Vogt:2010cv,LoPresti:2014ihe}
showed that the resummation of large threshold 
logarithms in off-diagonal channels has a non-trivial 
exponentiation pattern, even at LL accuracy, 
which involves the color structure $C_F - C_A$. 
The reason for this non-trivial behavior can be 
understood, for instance, by investigating the 
factorization structure of the Drell-Yan $qg$ 
channel by means of the method of regions,
which shows that both virtual collinear and 
soft modes contribute at LL accuracy.
Therefore, a simplification such as in 
Eq.~(\ref{sigmahadN2}) is not possible for 
off-diagonal channels. Despite this 
complication, it has been possible to
derive an all-order resummation formula 
for NLP LLs in off diagonal channels of 
processes such as DIS, Drell-Yan and Higgs
production \cite{Beneke:2020ibj,SCETunpublished,vanBeekveld:2021mxn}.
To this end, a fundamental input is provided
by deriving so-called consistency conditions,
obtained by requiring the cancellation of leading 
poles in the total cross section. One 
parameterizes the leading poles in the 
total cross section in terms of unknown 
coefficients multiplying the relevant 
scales of the process. Requiring pole 
cancellation in the bare total cross section 
provides a set of equations, whose solution 
determines the minimal number of parameters 
needed to fix all the poles -- and thus the 
related large logarithms. The derivation
is rather technical, and we refer to \cite{Beneke:2020ibj,vanBeekveld:2021mxn}
for further details. The relevant 
observation is that a minimal set of 
parameters is given entirely by those 
associated to a single scale 
in the problem -- one parameter for each 
perturbative order. It is thus possible
to obtain the entire cross section by
determining the full tower of coefficients 
associated, for instance, to the hard scale, 
or to the soft scale in the problem. 
Determining the all-order coefficients 
associated to the hard scale can be 
done by means of effective field theory 
methods, and we refer to \cite{Beneke:2020ibj} 
for further details. The diagrammatic approach,
instead, is particularly suitable to 
determine at all orders the coefficients 
associated to the leading poles in the 
soft region -- in a suitable gauge, 
it can be shown that the soft region 
is determined by ladder diagrams such as 
those represented in Fig.~\ref{fig:DYladder}
\cite{vanBeekveld:2021mxn}. 
These can be calculated to all orders,
and subsequently the consistency condition
allow one to fix the entire cross section. 
For instance, in the case of the $qg$ channel 
in Drell-Yan, the resummed NLP LL partonic 
cross section takes the form \cite{SCETunpublished,vanBeekveld:2021mxn}
\begin{eqnarray} \label{CDYgqres} \nonumber
\Delta_{qg}^{\rm dQCD,\,LP+NLP}
&=&\frac{T_R}{C_A-C_F}
\frac{1}{2N \ln N}\bigg\{
e^{\frac{2C_F \alpha_s}{\pi} \ln^2 N}
{\cal B}_0\bigg[\frac{\alpha_s}{\pi}(C_A-C_F)
\ln^2 N\bigg] \\ 
&&\hspace{4.0cm}-\, e^{(2C_F+6C_A)\frac{\alpha_s}{4\pi}\ln^2N}\bigg\},
\end{eqnarray}
where 
\begin{equation}
  {\cal B}_0(x)=\sum_{n=0}^\infty \frac{B_n}{(n!)^2}x^n\,,
  \label{B0def}
\end{equation}
and in turn $B_n$ are Bernoulli numbers, 
$B_0 = 1$, $B_1 = -1/2$, $\ldots$. Combining together 
Eq.~(\ref{eq:resumgeneral}) with
Eq.~(\ref{CDYgqres}) completes the 
resummation of LLs at NLP in all 
the production channels contributing 
to Drell-Yan. 
Analogous equations can be obtained
for Higgs boson production in gluon
fusion and DIS, and we refer to
\cite{Beneke:2020ibj,vanBeekveld:2021mxn}
for a detailed derivation.

\subsection{Effective field theory approach for cross-sections}

The effective field theory approach can achieve a proper, systematic treatment of factorization and resummation beyond the leading power. SCET  provides a framework to describe processes involving energetic and soft particles to all powers. It  allows for an operator-based expansion, maintaining homogeneous power-counting and gauge invariance at every step of the computations. 
The EFT formulation provides further insights into the structure of well-established soft theorems \cite{Larkoski:2014bxa,Beneke:2017mmf,Beneke:2021umj}, which are the basis for a systematic study of power suppressed effects. The EFT approach enables the study of subleading power terms in soft and collinear expansion not only in QCD but also in Gravity \cite{Beneke:2021aip} and it would be interesting to explore further applications. 

The SCET expansion parameter is typically denoted by $\lambda$. Its precise definition depends on the process under consideration, e.g., $\lambda\sim \sqrt{\xi}$ as in Eq. \eqref{threshold_simple} above. The theory contains collinear and soft modes. Each collinear direction has its collinear mode, i.e., a generic $N$-jet process will be described by $N$ collinear modes. Interactions within a given collinear sector and interactions with the soft background are contained in the Lagrangian. The complete Lagrangian can then be written as a sum over collinear directions 
\begin{align}
    \mathcal{L}_{\rm SCET} = \sum_{i=1}^{N} \mathcal{L}_i +\mathcal{L}_s,
\end{align}
where $\mathcal{L}_i$ is the $i$-collinear Lagrangian and  $\mathcal{L}_s$ is a pure soft Lagrangian. In principle, one also has to consider the Glauber Lagrangian \cite{Rothstein:2016bsq}. However, for many observables, the Glauber contribution cancels and thus this term is often ignored.  
The $i$-collinear Lagrangian can then be written as a series in $\lambda$
\begin{equation}
    \mathcal{L}_i = \mathcal{L}_i^{(0)} + \mathcal{L}_i^{(1)} + \mathcal{L}_i^{(2)} + \mathcal{O}(\lambda^3)\,.
\end{equation}

Interactions between different collinear sectors are obtained after integrating out the hard modes. They are encoded in the currents, which are constructed from the collinear gauge-invariant building blocks \cite{Bauer:2000yr,Bauer:2001ct}. 
At leading power, we can only use a single building block 
\begin{align}\label{eq:JA0Blocks}
 J_i^{A0}(t_i) \in \left\{\chi_i(t_i n_{i+}),\overline{\chi}_i(t_i n_{i+}),\mathcal{A}_{i\perp }(t_i n_{i+}) \right\}   ,
\end{align}
for each collinear direction. Here $\chi_i(x) = W_i^\dagger(x) \xi_i(x)$ is a fermionic building block, constructed out of the $i$-collinear field $\xi_i(x)$ and the collinear Wilson line $W_i$. Similarly, the gluon building block is related to the collinear gluon field. 
Since the building blocks scale as $\lambda$, adding more $i$-collinear fields in one direction creates power-suppressed operators. Alternatively, power suppression can be introduced by the action of a derivative on the collinear fields. Details on the  operatorial basis construction and renormalization of the subleading operators for various cases can be found in \cite{Moult:2015aoa,Kolodrubetz:2016uim,Beneke:2017ztn,Beneke:2017mmf,Moult:2017rpl,Chang:2017atu,Feige:2017zci,Beneke:2018rbh,Beneke:2019kgv,Moult:2019mog}.

While the study of power corrections in the context of flavor physics has a long history \cite{Beneke:2003pa,Hill:2004if,Lee:2004ja,Beneke:2004rc,Beneke:2004in,Paz:2009ut,Benzke:2010js,Beneke:2015wfa}, the first resummation using SCET for event shape-type observables was achieved relatively recently for thrust distribution \cite{Moult:2018jjd}. Subsequently, threshold resummation has been studied for Drell-Yan \cite{Beneke:2018gvs} and Higgs production \cite{Beneke:2019mua}.  Each of these studies considers only diagonal channels, i.e., not involving soft quarks and achieved leading logarithmic accuracy. 
The LL results have a relatively simple structure for all these processes. For example, the Drell-Yan threshold resummation leads to the following expression
\begin{align}\label{eq:DYNLP}
\Delta^{\rm LL}_{\rm NLP}(z,\mu) =  
\frac{\hat\sigma^{\rm LL}_{\rm NLP}(z,\mu)}{z} =&
\exp \left[ -2 \frac{\alpha_s C_F}{\pi}
\ln^2\frac{\mu}{\mu_h} \right]\,
\exp \left[+2 \frac{\alpha_s C_F}{\pi}
\ln^2\frac{\mu}{\mu_s}\right]\nonumber\\
& \times (-4) \frac{\alpha_s C_F}{\pi}
\ln\frac{\mu_s}{\mu}\,\theta(1-z)\,.
\end{align}
One can recognize the subleading power logarithm appearing at NLO in the second line, dressed by Sudakov exponents originating from the hard and soft functions. Interestingly, even these simple results required the development of new techniques allowing to perform the resummation. These are related to the observation that the leading logarithms appear due to operator mixing under renormalization \cite{Moult:2018jjd,Beneke:2018gvs}. 

\begin{figure}[t]
\begin{center}
  \includegraphics[width=0.45\textwidth]{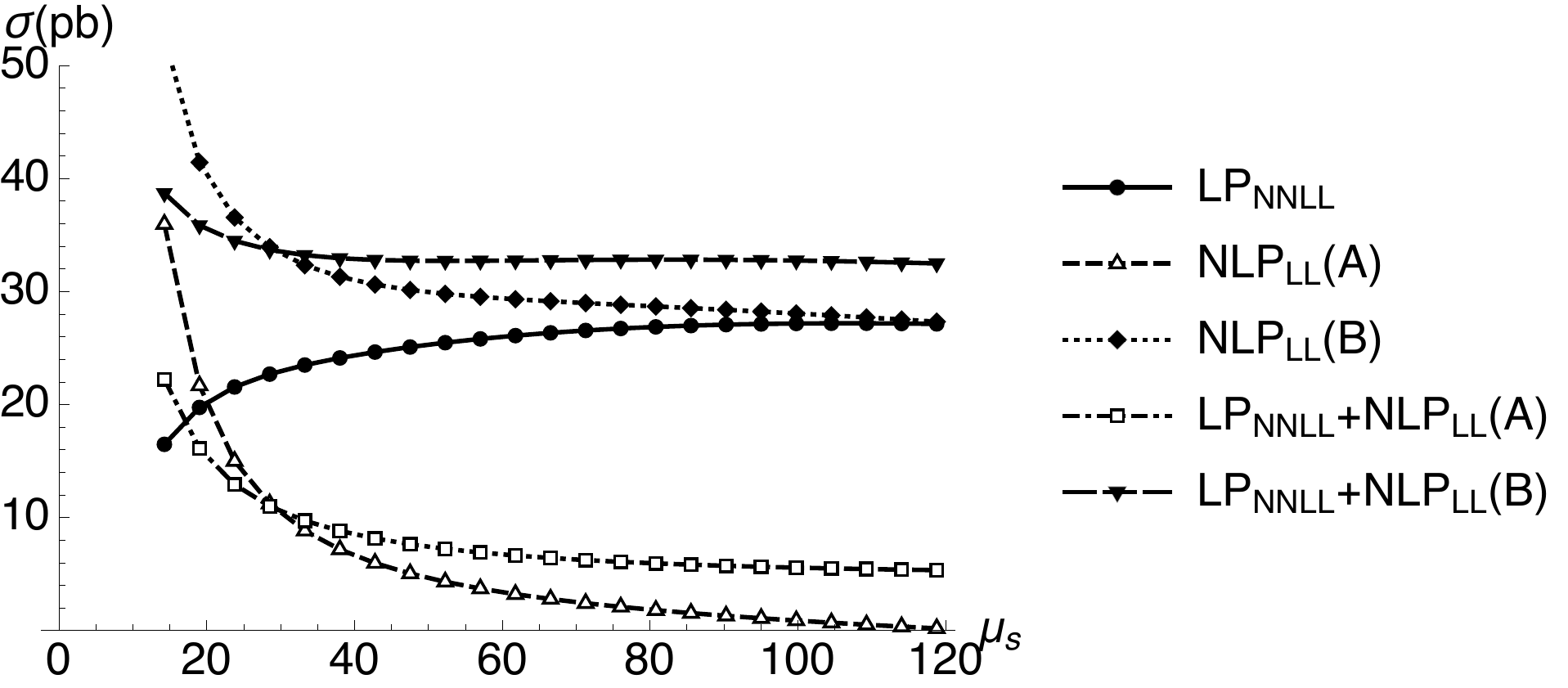}
  \qquad
  \includegraphics[width=0.45\textwidth]{sigmamusB}
\end{center} 
\caption{Soft scale dependence of the NNLL LP and LL NLP resummed Higgs production 
cross section. Left panel: $\mu_h^2 = m_H^2$, right panel: $\mu_h^2 = -m_H^2$.  Letters $A$ and $B$ refer to different possible choice of initial conditions, see  \cite{Beneke:2019mua}.}
\label{fig:sigmamus}
\end{figure}

The result for the Higgs threshold production \cite{Beneke:2019mua} can be obtained from the Drell-Yan result~(\ref{eq:DYNLP}) after substituting color factors $C_F \leftrightarrow C_A$. Phenomenological study for $13 \rm TeV$ proton-proton collisions shows that the LL NLP corrections can exceed 30\% of the NNLL LP resummed cross section. However, as shown in Fig.~\ref{fig:sigmamus}, the dependence on the soft scale is substantial and indicates a need for extension for the NLP resummation beyond LL accuracy, and possibly even further terms in the power expansion \cite{Anastasiou:2015vya}.

While the general procedure of deriving subleading power factorization theorems is well understood \cite{Beneke:2019oqx,Moult:2019mog}, the resummation beyond leading logarithmic accuracy for diagonal channels is still an open problem.  The so-called endpoint divergent problem hinders the extension of the above results beyond LL accuracy or to the off-diagonal contributions involving soft quarks. The factorization theorems at NLP generically are a sum over several terms, where each term is a convolution between the hard, collinear, and soft functions. While these convolution integrals are well defined for the bare factorization theorem in $d$-dimensions, the limit $d\to4$ does not exist as was explicitly demonstrated in \cite{Beneke:2019oqx,Jaskiewicz:2019phy}. This problem originates, for example, due to the overlap between soft and collinear modes. In \cite{Beneke:2019kgv}, it has been observed that for certain cases protected by reparameterization invariance, one can redefine the operator basis to remove the endpoint divergences. 
In \cite{Moult:2019uhz} the result for the Sudakov factor involving soft quarks has been conjectured and it has been shown that endpoint divergences are absent for $\mathcal{N}=1$ QCD, i.e. when the color charge of quark and gluon are equal. The conjectured soft quark Sudakov factor has been subsequently proven and generalized in \cite{Beneke:2020ibj}. 
This has been possible thanks to the development of the additional endpoint factorization of the divergent contributions and the introduction of new auxiliary modes, which parameterize the overlap between soft and collinear contributions. Using this result,  the short-distance coefficient in the $\overline{{\rm{MS}}}$ scheme in off-diagonal Higgs induced DIS can be obtained in closed form, yielding the following result 
\begin{eqnarray}
\tilde C_{\phi,q}^{NLP,LL}\Big|_{\epsilon\to 0}
&=& \frac{1}{2N\ln N}\frac{C_F}{C_F-C_A}\Bigg(
{\cal B}_0(a)\exp\left[C_A\frac{\alpha_s}{\pi}\left(\frac12\ln^2 N+\ln N\ln\frac{\mu^2}{Q^2}\right)\right]
\nonumber\\
&&-\, \exp\left[\frac{\alpha_s C_F}{\pi}\left(\frac12\ln^2 N+\ln N\ln\frac{\mu^2}{Q^2}\right)\right]\Bigg)\,,
\label{eq:Cqgfinite}
\end{eqnarray}
where $a = \frac{\alpha_s}{\pi} (C_F-C_A) 
\ln^2 N\,$ and $ \mathcal{B}_0(x) $ is defined in \eqref{B0def}. The above result
agrees with Eq.~(29) of \cite{Vogt:2010cv} for $\mu=Q$, 
and generalizes it to $\mu\not= Q$.

Dealing with the endpoint divergences is currently at the forefront of EFT studies. Once a systematic treatment is established, we can expect the accuracy of resummation to be improved. Not much is known about the properties of the subleading soft and jet functions. Further progress will require a better understanding of the renormalization properties of these complicated non-local objects. 
Beyond the applications of SCET$_{\rm I}$ that are discussed here, there are new interesting areas of study in SCET$_{\rm II}$ involving subleading power rapidity divergences for transverse momentum distributions~\cite{Ebert:2018gsn}, Energy-Energy correlators~\cite{Moult:2019vou}, Regge kinematics and forward scattering \cite{Moult:2017xpp,Bhattacharya:2021zsg}, and azimuthal asymmetries in semi-inclusive deep inelastic scattering \cite{Ebert:2021jhy}. 

Progress on the application of the EFT methods to power expansion for collider observables will be crucial for achieving the precision required by the High-Luminosity LHC, EIC and future colliders. Applications of a systematic power expansion and resummation with the help of the renormalization group equations will enable us to achieve better precision and has a potential to impact the development of subtraction methods for fixed order computations \cite{Moult:2016fqy,Boughezal:2016zws,Moult:2017jsg,Ebert:2018lzn,Bhattacharya:2018vph,Ebert:2018gsn,Boughezal:2018mvf,Boughezal:2019ggi}.

\subsection{Diagrammatic methods for mass-  and angle-suppressed amplitudes}

Two of the most phenomenologically and conceptually interesting
examples of power corrections are given  by the mass-suppressed
amplitudes in processes with massive fermions  in the
fixed-angle {\it Sudakov} limit and the amplitudes suppressed
by the ratio of a characteristic momentum transfer to the
total energy in the small-angle or {\em Regge} limit.

In processes with massive fermions,  the origin of the
logarithmic corrections and the asymptotic behavior of the
amplitudes at next-to-leading power (NLP) drastically differ from
the classical leading-power Sudakov result
\cite{Sudakov:1954sw}. The double-logarithmic terms in this
case are related to the effect of the eikonal color charge
nonconservation in the process with soft fermion exchange and
result in asymptotic exponential enhancement for a wide class
of amplitudes and a breakdown of a formal power counting
\cite{Penin:2014msa,Liu:2017vkm}. 

To illustrate the general properties of such processes, we
consider the amplitude of light quark  mediated Higgs boson
production in gluon fusion which is suppressed by the quark to
Higgs boson mass ratio $m_q^2/m_H^2$ relative to the leading
contribution of the top quark loop. With the targeted precision
of the QCD prediction for the cross section in  a percent range
\cite{Anastasiou:2016cez} the  contributions of bottom quark
loop cannot be ignored and the  problem of its theoretical
description ultimately goes beyond the finite order
perturbative calculation \cite{Lindert:2017pky,Czakon:2020vql}.
Indeed, the radiative corrections in this case are enhanced by
the second  power of the large logarithm $\ln(m_H/m_q)$ with
the effective expansion parameter being
$\alpha_s\ln^{2}(m_H/m_b)\approx 40\alpha_s$. In the leading
(double) logarithmic approximation, the light quark mediated $gg
\to H$ amplitude  reads \cite{Liu:2017vkm,Liu:2018czl}
\begin{equation}
{{\cal M}^q}^{LL}_{gg\to H}=
-{\frac{3}{2}}{Z^2_{g}} g(z){\frac{m_q^2}{m_H^2}}
\ln^2\left({\frac{m_H^2}{m_q^2}}\right)
\,{{\cal M}^{t}}^{(0)}_{gg\to H}\,,
\label{eq::MqLL}
\end{equation}
where ${{\cal M}^{t}}^{(0)}_{gg\to H}$ is the leading
order heavy top quark mediated  amplitude.
Eq.~(\ref{eq::MqLL}) reveals a number of  characteristic
features of mass-suppressed processes. It includes the
standard infrared divergent Sudakov factor $Z^2_{g}$ for the
external on-shell gluon  states and the function $g(z)$ of the
variable $z=(C_A-C_F){\alpha_s\over 4\pi}\ln^2(m_H^2/m_q^2)$
which accounts for the non-Sudakov mass logarithms to all
orders in $\alpha_s$. It can be solved in terms of the
generalized hypergeometric function
$g(z)={}_2F_2\left(1,1;{3/2},2;{z/2}\right)$ and has  the
following asymptotic behavior at $z\to\infty$
\begin{equation}
g(z)\sim \left({2\pi e^{z}\over z^{3}}\right)^{1/2}\!\!,
\label{eq::gasymp}
\end{equation}
i.e. in contrast to the Sudakov factor it results in the
exponential enhancement of the amplitude in the high-energy
(small quark mass) limit. The difference $C_A-C_F$ of the
quadratic Casimir operators of the gluon and quark color group
representations  in the definition of the variable $z$ accounts
for the color charge  variation  of the highly energetic parton
after emission of a soft quark, which constitutes  the physical
origin of the non-Sudakov double logarithms.

Though the leading logarithmic result determines the
qualitative behavior of the amplitude, it cannot be used for a
reliable quantitative estimates since it does not account for
the numerically important effects such as the renormalization
scale dependence of the quark mass and coupling constants,
which requires the inclusion  of subleading logarithms. In
Ref.~\cite{Anastasiou:2020vkr}, the next-to-leading logarithmic
corrections to Eq.~(\ref{eq::MqLL}) of  the form
$\alpha_s^n\ln^{2n-1}(m_H/m_q)$ for all $n$ have been computed
and applied to the analysis of the Higgs boson production in
threshold approximation.  For the yet unknown
next-to-next-to-leading and next-to-next-to-next-to-leading
bottom  quark contribution to the total cross section this
analysis gives $-0.12~pb$ and $-0.02~pb$, respectively. With a
rather conservative assessment of the accuracy of the
next-to-leading logarithmic and the threshold approximations it
provides a rough estimate of the bottom quark mediated
contribution to the total cross section of Higgs boson
production in gluon fusion beyond NLO to be in the range from
$-0.32$ to $0.08~pb$, thereby reducing the previous uncertainty
estimate  \cite{Anastasiou:2016cez} by a factor of two.

The actual  accuracy of the  logarithmic and threshold
approximations, however,  is difficult to estimate, and the
above interval has to be further  reduced by evaluating  the
next-to-next-to-leading logarithmic contribution and getting an
approximation valid beyond the threshold region. The latter
requires the analysis of the logarithmically enhanced
corrections to the hard real emission which is currently not
available even in the leading logarithmic approximation (only
the abelian part of the double-logarithmic corrections  for the
$gg \to Hg$ amplitude of Higgs plus jet production has been
obtained in Ref.~\cite{Melnikov:2016emg}). The extension of the
method \cite{Liu:2017vkm,Liu:2018czl,Anastasiou:2020vkr} beyond
the next-to-leading logarithms and to processes with hard
real radiation is one of  big challenges for the modern
effective field theory and is of primary phenomenological
importance for the high-precision Higgs physics program at the
LHC.  At the same time the high-order next-to-next-to-leading power terms are well under control \cite{Liu:2021chn}.

The Regge limit  of high-energy scattering describes a
kinematical configuration with vanishing  ratio of the
characteristic momentum transfer to the total energy of the
process.  The relevant expansion parameter is given by the
ratio of the Mandelstam variables $t/s$. Despite a crucial
simplification due to decoupling of the light-cone and
transversal degrees of freedom, the gauge interactions in this
limit possess highly nontrivial dynamics giving a rigorous
quantum field theory realization of the Regge concept for
high-energy scattering. Major progress has been achieved in the
analysis of the leading-power amplitudes which  culminated in
the evaluation  of the next-to-leading QCD corrections to  the
theory of  BFKL pomeron \cite{Fadin:1998py}.

The expansion in $t/s$ is also mandatory for the evaluation of
the nonfactorizable QCD corrections to the Higgs boson
production via vector boson fusion. The leading order of
this expansion is equivalent to the eikonal  approximation,
which reduces the problem to the effective theory analysis in
the two-dimensional transversal space. This analysis has been
performed at next-to-next-to-leading order in QCD for the
cross sections of a single Higgs boson \cite{Liu:2019tuy} and
Higgs boson pair production   \cite{Dreyer:2020urf}.

At the same time, very little is known about the structure of
the logarithmic corrections and the  asymptotic behavior of the
power suppressed amplitudes which are crucial to control the
accuracy and validity of the analysis based on the eikonal
approximation. Only recently the leading logarithmic result for
the power-suppressed part of the fermion scattering amplitude
has been obtained in an abelian gauge theory within an
effective field theory framework \cite{Penin:2019xql}. In
contrast to the single-logarithmic Regge behavior of the
leading-power amplitude, the radiative corrections to the
power-suppressed term exhibit a double-logarithmic  enhancement
and the result for the amplitude has the following form
\begin{equation}
{\cal M}^{LL}=|t/s| z\,g(2z){\cal M}^{\rm Born}\,,
\label{eq::llamp}
\end{equation}
where $z={\alpha\over 2\pi}\ln(|t/s|)$ and  $g(z)$ is,
amazingly, the same function as in Eq.~(\ref{eq::MqLL})
describing the non-Sudakov corrections to the mass suppressed
amplitude. Due to the exponential growth of this function at
large values of its argument, at  $\left|{s/t}\right|\approx
e^{2\pi/\alpha}$, the formally power-suppressed contribution to
the scattering cross section becomes comparable to the
leading-power result and the small-angle expansion breaks down.
This, in particular, may be considered as a possible solution
of the long standing unitarity problem of the Regge analysis of
the leading-power scattering amplitudes in QED
\cite{Frolov:1970ij}. The generalization of the above result to the
nonabelian  gauge groups and more complex amplitudes is
therefore of primary theoretical interest.

\subsection{Effective field theory approach for mass suppressed amplitudes}

In parallel to the direct QCD approach, the mass suppressed effects on the amplitude level have been studied in the SCET$_{\rm II}$ framework. It has been shown that these effects give rise to unexpectedly large QED corrections for the leptonic decays of $B_s$ mesons \cite{Beneke:2017vpq}. They have been resummed, in part, in \cite{Beneke:2019slt}. The resummed corrections were due to a contribution which is not endpoint divergent thanks to the stronger suppression of the endpoint region at the hard scale. 

The breakthrough has been achieved in the study of the $h\to \gamma \gamma $ amplitude mediated by the light quark. The condition on cancellation of the rapidity divergences has been used to derive refactorization conditions \cite{Liu:2019oav} and perform the LL resummation. The renormalization properties of the NLP jet and soft functions have been subsequently studied in \cite{Liu:2020eqe,Bodwin:2021cpx,Liu:2020ydl,Liu:2021mac}. These objects have a close connection to functions appearing earlier in the flavor applications of SCET. Generally, they are simpler than the corresponding objects defined at the level of the cross section in SCET$_{\rm I}$, as their dependence on the convolution variables can be traced back to a particular external state which defines the amplitude.

The refactorization conditions were then rigorously derived in \cite{Liu:2020wbn} as operatorial identities. This allowed proving that the reshuffling of the factorization theorem removes endpoint divergences. Using the previously computed anomalous dimensions of soft and jet functions led to the first complete NLL resummation \cite{Liu:2020tzd} for mass suppressed effects. 

Resummation of power suppressed amplitudes will have a substantial impact on the precision in studies of the top and Higgs sectors at the LHC. Applications of the EFT methods will  impact also flavor studies and in particular the evaluation of the QED corrections, which are a necessary ingredient for the resolution of flavor anomalies.  

\section{Jets and their substructure}
\label{sec:jets}

\begin{figure}[t]
\begin{center}
\includegraphics[width=0.49\textwidth]{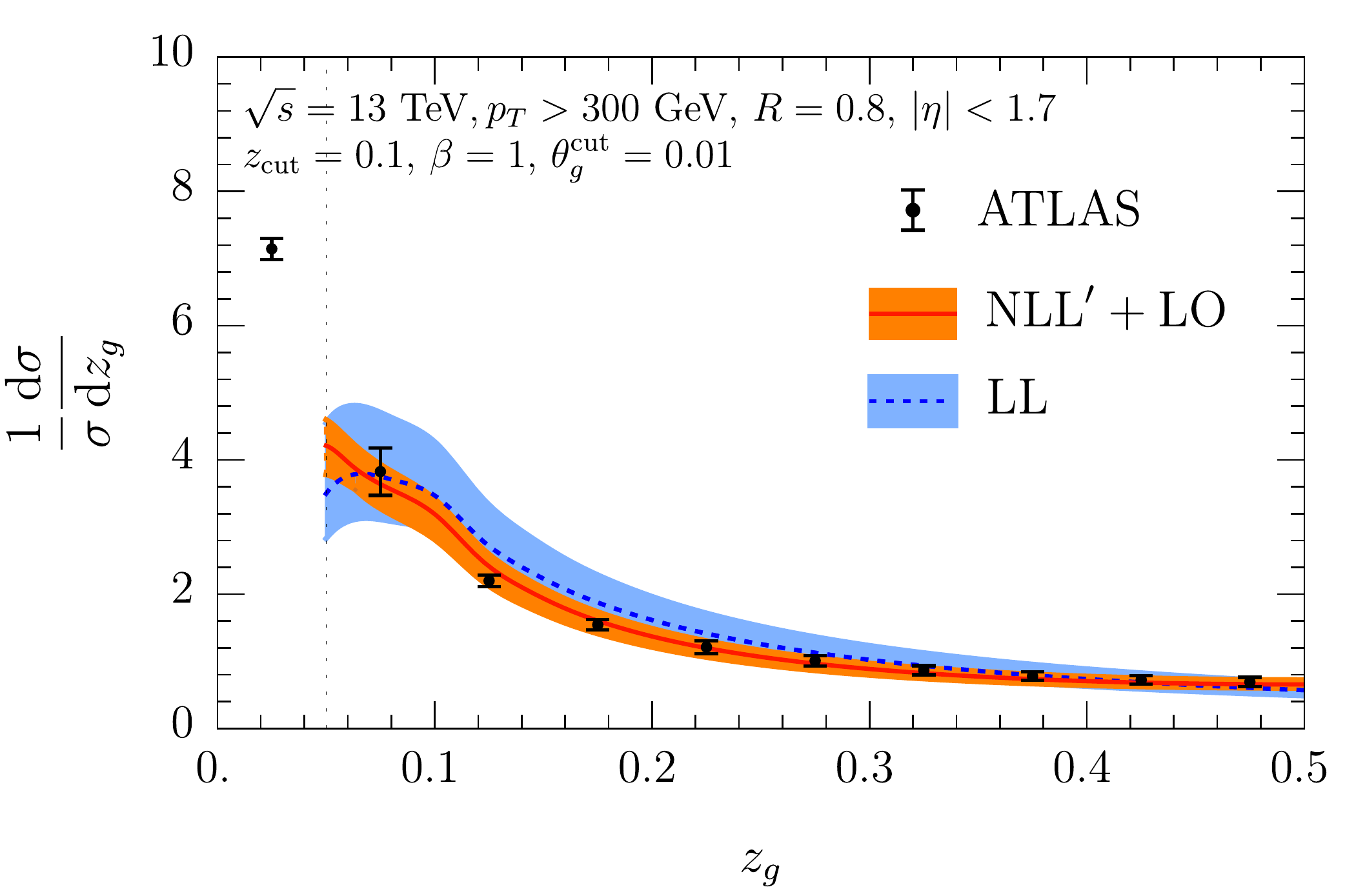}
\includegraphics[width=0.49\textwidth]{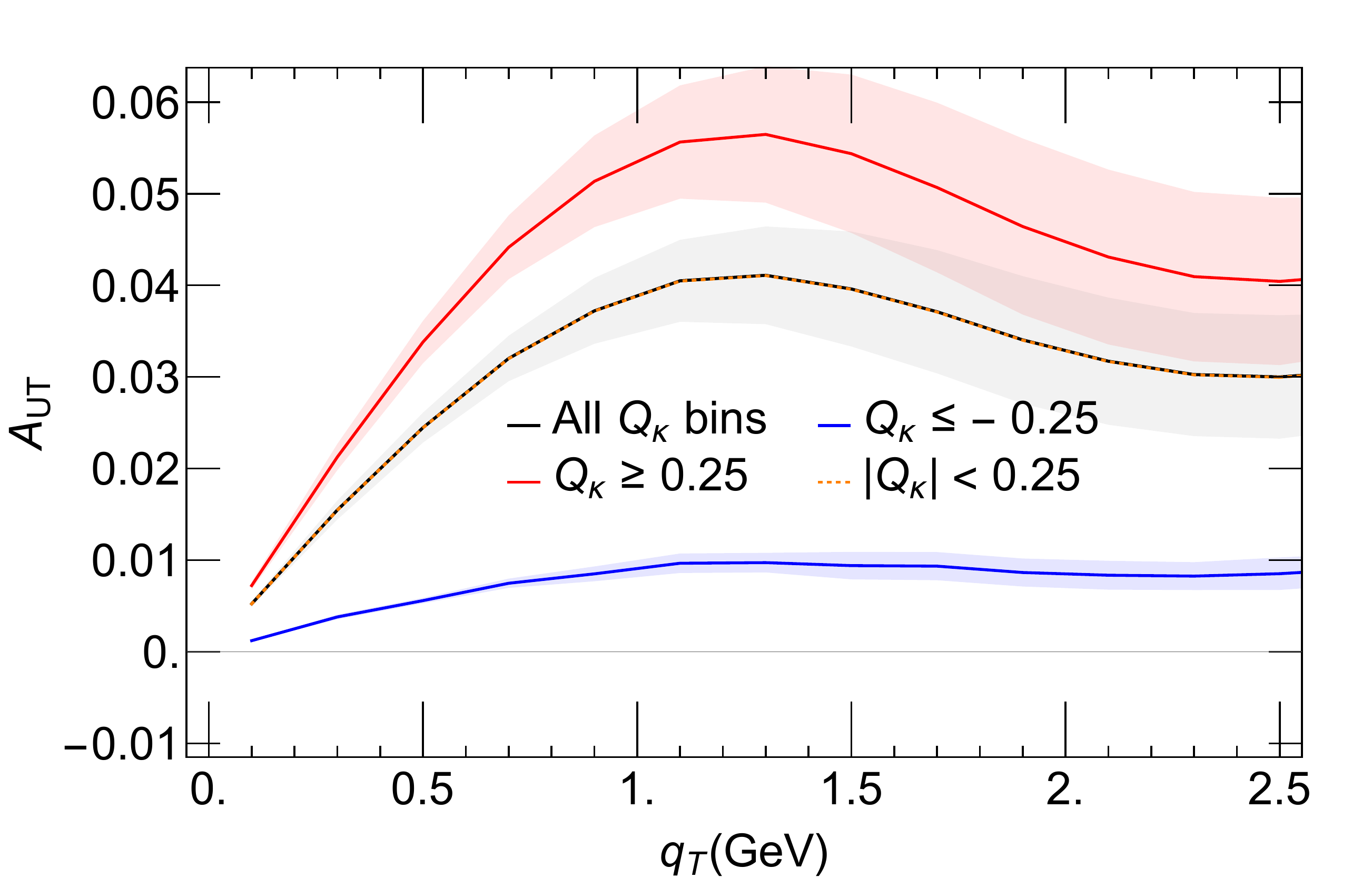}
\caption{Left: Comparison of numerical results for the momentum sharing fraction $z_g$ at NLL$'$ accuracy~\cite{Cal:2021fla} to ATLAS data~\cite{ATLAS:2019mgf}. Figure taken from Ref.~\cite{Cal:2021fla}. Right: Prediction of the Sivers asymmetry at the EIC in back-to-back electron-jet production for different bins of the jet charge $Q_\kappa$. Figure taken from Ref.~\cite{Kang:2020fka}~\label{fig:JSS}}
\end{center}
\end{figure}

Jets are collimated sprays of particles which are observed in the detectors of high-energy scattering experiments such as the LHC, RHIC, LEP, HERA and the future EIC. They reflect the underlying quark and gluon degrees of freedom which emit radiation approximately collinear to their initial direction. In recent years, significant progress has been made in improving the perturbative precision of perturbative QCD calculations. For example, fixed order calculations for inclusive jet production are now available at NNLO~\cite{Currie:2016bfm,Czakon:2019tmo}. In addition, logarithmic corrections at threshold and in the jet radius have been resummed at NLL$'$ accuracy~\cite{Dasgupta:2014yra,Liu:2017pbb,Neill:2021std}. In the future, it will be critical to extend these calculations to N$^3$LO and next-to-next-to leading logarithmic (NNLL) accuracy in order to match the experimental precision. In particular, the field of jet substructure has received a growing attention over the last decade~\cite{Larkoski:2017jix,Kogler:2018hem,Marzani:2019hun}. Here the goal is to characterize and utilize the radiation pattern inside jets. Jet substructure observables are relevant for searches of BSM physics and various precision measurements of Standard Model processes such as Transverse Momentum Dependent (TMD) PDFs and fragmentation functions~\cite{Procura:2009vm,Bain:2016rrv,Kang:2017glf,Neill:2016vbi,Neill:2018wtk,Gutierrez-Reyes:2019msa}, the strong coupling constant $\alpha_s$~\cite{Proceedings:2018jsb,Marzani:2019evv}, quarkonia~\cite{Bain:2017wvk,Kang:2017yde}, and the top quark mass~\cite{Hoang:2017kmk} and to extract medium properties in heavy-ion collisions. The need for precision calculations will increase during the high-luminosity era of the LHC, where the detectors will be optimized for the first time to perform precision jet substructure measurements. In addition, the EIC is expected to perform precise measurements of low-energy jets. The clean environment of the EIC will allow for novel measurements complimentary to the LHC. In addition, RHIC and LEP data are analyzed to measure jet energy spectra and jet substructure observables. Jet grooming techniques~\cite{Krohn:2009th,Ellis:2009me,Dasgupta:2013ihk,Cacciari:2014gra,Larkoski:2014wba} have been developed to systematically remove soft radiation from identified jets which can be difficult to account for from first principles in QCD. These techniques can reduce nonperturbative effects and allow for direct and precise comparisons between theoretical calculations and data. Jet substructure observables at collider experiments typically involve the measurement of multi-differential cross sections. This necessitates the joint resummation of several large logarithmic corrections to all orders in the QCD strong coupling constant, which is the main theoretical challenge. Examples include logarithms in the jet mass, jet radius and grooming parameters. These different large logarithmic corrections are generally not independent and require a careful identification of relevant energy scales and corresponding factorization theorems. Depending on the relative scaling of the involved variables, different factorization theorems need to be derived. Eventually, the different results need to be merged to allow for a meaningful comparison to experimental data. Moreover, observables like the groomed momentum sharing fraction $z_g$ are Sudakov safe~\cite{Larkoski:2015lea} and require, in addition, the joint resummation of an auxiliary variable like the groomed jet radius $\theta_g$. To illustrate the complexity of factorization theorems for jet substructure observables, we consider the momentum sharing fraction $z_g$~\cite{Larkoski:2015lea}. It is of considerable phenomenological interest as it allows for the most direct measurement of the QCD splitting function. For inclusive jet production, jet functions ${\cal G}_{i=q,g}$ that depend on the jet substructure measurement can be factored out, which are then multiplied by appropriate quark/gluon fractions. In order to resum all relevant large logarithmic corrections, the jet functions needs to be refactorized as~\cite{Cal:2021fla}
\begin{align}\label{eq:zg}
   {\cal G}_i 
    &= \Theta(1/2>z_g>z_{\rm cut}\theta_g^\beta)\,
\tilde H_i(p_TR,\mu)\,C_i^{\in {\rm gr}}(\theta_g p_T R,\mu)
\, S_i^{\notin {\rm gr}}(z_{\rm cut}p_T R,\beta,\mu)\,
    \nonumber \\ & \quad 
    \tilde {\cal S}_G(z_{\rm cut}\theta_g^{1+\beta}p_T R,\beta,\mu)\,    
   S_i^{\rm NG}(z_{\rm cut}) \bigg[\frac{\rm d}{{\rm d}z_g} \frac{\rm d}{{\rm d}\theta_g} 
    \tilde {\cal S}_{z_g}(z_g\theta_g p_T R,\mu) 
    \nonumber \\ & \quad 
    + \tilde {\cal S}'^{\rm NG}_{i,1}(z_g \theta_g, z_g) 
    + \tilde {\cal S}'^{\rm NG}_{i,2}\Big(z_g \theta_g, \frac{z_g}{z_{\rm cut}\theta_g^\beta}\Big)  \bigg]\,.
\end{align}
Here, $z_{\rm cut},\beta$ denote parameters of the soft drop grooming procedure and the functions $\tilde {\cal S}'^{\rm NG}_{i,j}$ account for non-global contributions. The factorization in Eq.~(\ref{eq:zg}) enables the joint resummation of four classes of large logarithmic corrections in the jet radius $R$, the grooming parameter $z_{\rm cut}$ and the variables $z_g,\theta_g$. Currently, the achieved precision using the factorization in Eq.~(\ref{eq:zg}) is NLL$'$ accuracy, which can be extended to yet higher accuracy in the future. These results allow percent level comparisons to the available data without the need to model nonperturbative effects. As an example, we show the comparison to ATLAS data~\cite{ATLAS:2019mgf} in the left panel of Fig.~\ref{fig:JSS}.

In recent years, significant progress has been made in developing factorization formulas and improving the theoretical precision of theoretical calculations using both direct perturbative QCD methods and SCET. Several hallmark observables in the field of jet substructure at the LHC have now been extended systematically to NLL$'$ and partially NNLL accuracy. Examples include the jet mass~\cite{Frye:2016aiz,Marzani:2017mva,Kang:2018vgn,Larkoski:2020wgx,Benkendorfer:2021unv}, jet angularities~\cite{Ellis:2010rwa,Larkoski:2014uqa,Kang:2018vgn,Kang:2018qra,Caletti:2021oor,Reichelt:2021svh}, the groomed jet radius~\cite{Kang:2019prh}, the momentum sharing fraction $z_g$~\cite{Larkoski:2015lea,Cal:2021fla}, the jet shape~\cite{Cal:2019hjc}, the angle between different jet axes~\cite{Cal:2019gxa}, the jet pull~\cite{Larkoski:2019urm,Larkoski:2019urm,Bao:2019usu}, and the primary Lund plane~\cite{Lifson:2020gua}. A new jet grooming technique called dynamical grooming has been developed in Ref.~\cite{Mehtar-Tani:2019rrk,Caucal:2021bae}. Dedicated observables to distinguish boosted Higgs, $Z$, $W$ or BSM jets have been studied analytically in Refs.~\cite{Larkoski:2017cqq,Napoletano:2018ohv,Cavallini:2021vot}. These observables make use of the multi-prong structure of jets originating from boosted resonances compared to single-prong QCD jets. In Refs.~\cite{Chen:2020vvp,Chen:2020adz,Li:2021zcf}, energy-energy correlators in the collinear limit were discussed with are particularly suitable for extensions to higher accuracy. Recently, jet tagging has also been proposed as a useful tool to constrain collinear PDFs at the LHC~\cite{Caletti:2021ysv}. While jet grooming has been introduced initially to remove soft wide-angle radiation from the jet, it has recently been utilized to design Infrared-Collinear Safe observables which are particularly sensitive to soft radiation~\cite{Chien:2019osu,Cal:2019gxa}. These observables may lead to a better understanding of the underlying event contribution and hadronization effects. In the future it will be essential to extend these calculations to NNLL$'$ or even N$^3$LL accuracy. This requires progress in fixed order calculations of relevant collinear and soft functions. For example, in Ref.~\cite{Liu:2021xzi}, the exclusive collinear jet function for anti-$k_T$  jets~\cite{Cacciari:2008gp} was computed at NNLO which is an essential ingredient to extend the precision of existing jet substructure calculations. See also Refs.~\cite{Kardos:2020ppl,Kardos:2020gty}. Moreover, future developments are necessary to resum non-global logarithms. In Ref.~\cite{Banfi:2021owj}, the NLL resummation of NGLs was achieved for the $e^+e^-$ hemisphere case which sets the starting point for extending the precision of relevant jet substructure observables.

At the EIC, precision calculations of jets and their substructure will be critical to explore the 3D structure of nucleons and nuclei and study aspect of the QCD hadronization process. Inclusive jet production has been calculated up to NNLO in unpolarized and polarized $ep$ scattering which can provide important constraints for collinear PDFs. See Refs.~\cite{Hinderer:2015hra,Abelof:2016pby,Boughezal:2018azh,Borsa:2020yxh,Borsa:2021afb} for recent developments. Jets are expected to play an important role to constrain TMD PDFs and fragmentation functions. Different than Semi-Inclusive Deep-Inelastic Scattering (SIDIS), jet observables allow for a decorrelation of initial and final-state TMD dynamics. Back-to-back electron-jet correlations measured in the laboratory frame~\cite{Liu:2018trl} can be used to study TMD PDFs which is independent of TMD fragmentation. In Ref.~\cite{Kang:2020fka}, it was proposed to use the jet charge~\cite{Field:1977fa,Krohn:2012fg} to separate up and down quark contributions to the Sivers asymmetry in transversely polarized $ep$ scattering. The numerical predictions for the EIC are shown in the right panel of Fig.~\ref{fig:JSS}. In the Breit frame, jets can be clustered in terms of energy instead of transverse energy or momentum as in the laboratory frame~\cite{Arratia:2019vju}. Since the photon virtuality $Q$ provides an additional reference scale, the jet energy spectrum can be accessed. This provides a unique opportunity to not only measure inclusive jets, but also the energy loss of leading jets in vacuum and the nuclear environment~\cite{Neill:2021std}. Breit frame jets also allow for measurements which are sensitive to TMD PDFs~\cite{Gutierrez-Reyes:2018qez}. Chiral-odd TMDs such as the transversity can be probed by time-reversal odd jets~\cite{Liu:2021ewb}. The observable and factorization structure is similar to SIDIS but the final-state TMD fragmentation function is replaced with a jet function that can be calculated perturbatively. Moreover, gluon TMD PDFs can be accessed using high transverse momentum di-jets in the Breit frame~\cite{delCastillo:2021znl,Kang:2021ffh}. TMD fragmentation can be studied independent of TMD PDFs using jet substructure observables. This includes both the unpolarized case and for example the Collins effect which can be accessed when the initial proton is transversely polarized~\cite{Yuan:2007nd,Arratia:2020nxw,Kang:2021ffh}. Energy-energy correlators in the back-to-back region can also provide important new insights into TMD dynamics~\cite{Gao:2019ojf}. Moreover, EIC jets provide a powerful tool to probe cold nuclear matter effects in electron-nucleus collisions~\cite{Liu:2018trl,Sievert:2018imd,Barata:2020rdn,Li:2020rqj,Zhang:2021tcc}. In order to match the precision of traditional processes such as SIDIS, higher perturbative accuracy is needed at leading and subleading power. 

\section{Small-$x$ resummation}
\label{sec:smallx}

Gluon saturation has attracted a lot of attention in recent years and is one of the major scientific goals of the future EIC~\cite{Accardi:2012qut}. Gluon saturation plays the key role in understanding proton and heavy nuclei collisions in the high-energy limit, where the gluon momentum fraction $x$ is very small. In this  small-$x$ region, the gluon density grows dramatically and enters the nonlinear regime where gluon recombination becomes equally important to the splitting process, and the color glass condensate (CGC) effective theory~\cite{McLerran:1993ni,McLerran:1993ka,McLerran:1994vd,Gelis:2010nm} is the proper framework to describe this regime. The nonlinear B-JIMWLK (or its infinite color approximation, the BK equation) equation~\cite{Mueller:1989st,Mueller:1993rr,Balitsky:1995ub, Kovchegov:1999yj, Jalilian-Marian:1997qno, Jalilian-Marian:1997jhx, Iancu:2000hn, Ferreiro:2001qy} replaces the linear parton evolution equations~\cite{Balitsky:1978ic}, which inevitably leads to gluon saturation~\cite{Gribov:1983ivg,Mueller:1985wy} with a characteristic scale $Q_s$. The saturation scale $Q_s$ features the typical transverse momentum of the gluons inside the proton or the nucleus and grows as $x$ decreases. 

There are experimental results~\cite{Golec-Biernat:1998zce,BRAHMS:2004xry,PHENIX:2004nzn,STAR:2006dgg,PHENIX:2005veb,Braidot:2010ig,ALICE:2018vhm,Morreale:2021pnn} that are compatible with saturation-model predictions, but there still exists no definitive evidence to claim a discovery. In the future, dedicated measurements at the EIC will provide further information on the regime of gluon saturation. 

In order to faithfully and unambiguously establish gluon saturation and its onset, reliable theoretical predictions for small-$x$ collider phenomena are crucial.
In the small-$x$ regime, the typical semi-hard saturation scale is of the order of a few GeV and, therefore, $\alpha_s(Q_s)$ is typically not small enough. As a consequence, calculations beyond the leading order (LO) are generally required to ensure the convergence of the perturbative results. Recently, some attempts have been made in realizing the next-to-leading order (NLO) calculations for small-$x$ physics~\cite{Dumitru:2005gt,Altinoluk:2011qy,Chirilli:2011km,Beuf:2011xd,Beuf:2016wdz,Beuf:2017bpd,Boussarie:2016ogo,Boussarie:2016bkq,Balitsky:2012bs,Roy:2018jxq,Liu:2019iml,Roy:2019cux,Roy:2019hwr,Caucal:2021ent}. However, systematic calculations remain challenging and often cut-offs need to be introduced that violate the small-$x$ factorization theorem and mix non-perturbative and perturbative quantities 
that can invalidate the predictions. Meanwhile, there also lacks a systematic way to resum the large logarithms in the semi-hard NLO coefficient of the small-$x$ factorization theorem, which leads to the infamous negative cross section problem~\cite{Stasto:2013cha} and hinders practical phenomenological applications of the NLO calculations. 

One recent proposal suggests to use power counting arguments and to couple SCET to the CGC formalism as a possible approach to combine systematic fixed order results and resummation~\cite{Kang:2019ysm, Liu:2020mpy}. 
The idea is applied to single-inclusive hadron production in proton-nucleus collisions, $pA\to hX$, at forward rapidity. The CGC factorization for this observable is confirmed at NLO~\cite{Chirilli:2011km}. However, the observed negative cross section when the hadron transverse momentum $p_{h,\perp}$ becomes a bit larger has been quite puzzling to the community~\cite{Stasto:2013cha}. The NLO partonic cross section contains contributions of the form
\bea\label{eq:NLO-thre}
& \frac{\mathrm{d}^2\hat{\sigma}^{(1)} }{\mathrm{d} z \mathrm{d}^2p_\perp'}
\propto - \frac{\alpha_s}{2\pi} {\bf T}^2_i   P_{i\to i}(z) \ln \frac{r_\perp^2\mu^2}{c_0^2}
  \left(1+ \frac{1}{z^2} e^{i \frac{1-z}{z}p_\perp'\cdot r_\perp} \right) \, \nn
& - \frac{\alpha_s}{\pi} {\bf T}_i^a {\bf T}_j^{a'}  
\int \frac{\mathrm{d}x_\perp}{\, \pi}
\left\{ \frac{1}{z} \frac{2}{(1-z)_+} 
\,  e^{i\frac{1-z}{z} p_\perp'\cdot r_\perp'} 
\frac{r_\perp' \cdot r_\perp''}{{r_\perp'}^2 {r_\perp''}^2} \right.  \nn 
& \hspace{2.ex} \left. +  \,  \delta(1-z) 
\ln \frac{X_f}{X_A}
\,
\left[ \frac{r_\perp^2}{{r_\perp'}^2 {r_\perp''}^2} \right]_+  \right\} W_{aa'}(x_\perp) 
+ \dots \,,
\eea
where $p_\perp'$ is the transverse momentum of the fragmenting parton. Here, $W_{aa'}$ is the CGC Wilson line in the adjoint representation, $X_A$ is the momentum fraction carried by the gluon from the nucleus and $X_f$ is the scale due to the rapidity divergence~\cite{Fleming:2014rea,Rothstein:2016bsq,Liu:2019iml,Kang:2019ysm}. Moreover, $P_{i \to i}(z)$ is the QCD splitting function. Other conventions of the notation can be found in~\cite{Liu:2020mpy}. The terms in the second line are due to the interference between initial and final state radiation. At forward rapidity, the interference is not power suppressed. It was explicitly demonstrated in~\cite{Liu:2020mpy} that as the hadron transverse momentum $p_{h,\perp}$ increases, $z$ quickly approaches $1$, and the threshold logarithms of the form $\frac{1}{(1-z)_+}$ overwhelmingly dominate over the other contributions and drive the cross section to negative values. Therefore, an appropriate threshold resummation is required to ensure a reliable prediction.

It is worthwhile to point out that the threshold resummation in the small-$x$ regime is different from the threshold resummation within the collinear factorization, as is obvious from Eq.~(\ref{eq:NLO-thre}) where the color structure ${\bf T}_i^a {\bf T}_j^{a'} W_{aa'}$ of the single logarithmic interference term proposes new challenges and can hardly be produced by the known threshold resummation techniques. Numerically, it is found that it is this interference term that makes up the dominant bulk of the threshold contribution~\cite{Liu:2020mpy}. Therefore, a new resummation formalism is required.

In~\cite{Liu:2020mpy}, the couplings between the CGC Wilson line $W_{aa'}(x)$ and the SCET soft and collinear fields are put in by hand. Invoking SCET power counting and factorization, it is shown that the cross section for this process can be factorized as 
\bea\label{eq:forward-LL}
&\frac{ \mathrm{d}^2\sigma  }{\mathrm{d}y_h \mathrm{d}^2p_{h\perp} }
=
   \sum_{i, j = g,q}
   \frac{1}{4\pi^2} \int \frac{ \mathrm{d}  \xi }{\xi^2} \,
\frac{\mathrm{d} x}{x}\,
z x f_{i/P}(x,\mu) D_{h/j}(\xi,\mu) \nn 
& \times  \int \mathrm{d}^2 b_\perp \mathrm{d}^2 b_\perp' 
 \, 
e^{ip_\perp' \cdot r_\perp } \,
  \Big \langle \langle 
   {\bf {\cal M}}_0 (b_\perp') | \, 
\bm{{\cal J}}(z,\mu,\nu,b_\perp,b_\perp')  
\bm{{\cal S}}(\mu,\nu,b_\perp,b_\perp')|{\bf {\cal M}}_0 (b_\perp) 
\rangle 
\Big \rangle_\nu  \,.  
\eea
where $\bm{{\cal J}}$ and $\bm{{\cal S}}$ represent the collinear and soft contributions, respectively. The LO color space notation 
$|{\cal M}_0(b_\perp) \rangle$~\cite{Catani:1999ss} is used and encodes the CGC (Glauber) Wilson line
$W_{i_cj_c}(b_\perp)$ where $i_c$ and $j_c$ are the color indices for the incoming and the outgoing partons using the fundamental for quarks and the adjoint representation for gluons. 
\begin{figure}[t]
\begin{center}
\includegraphics[width=0.49\textwidth]{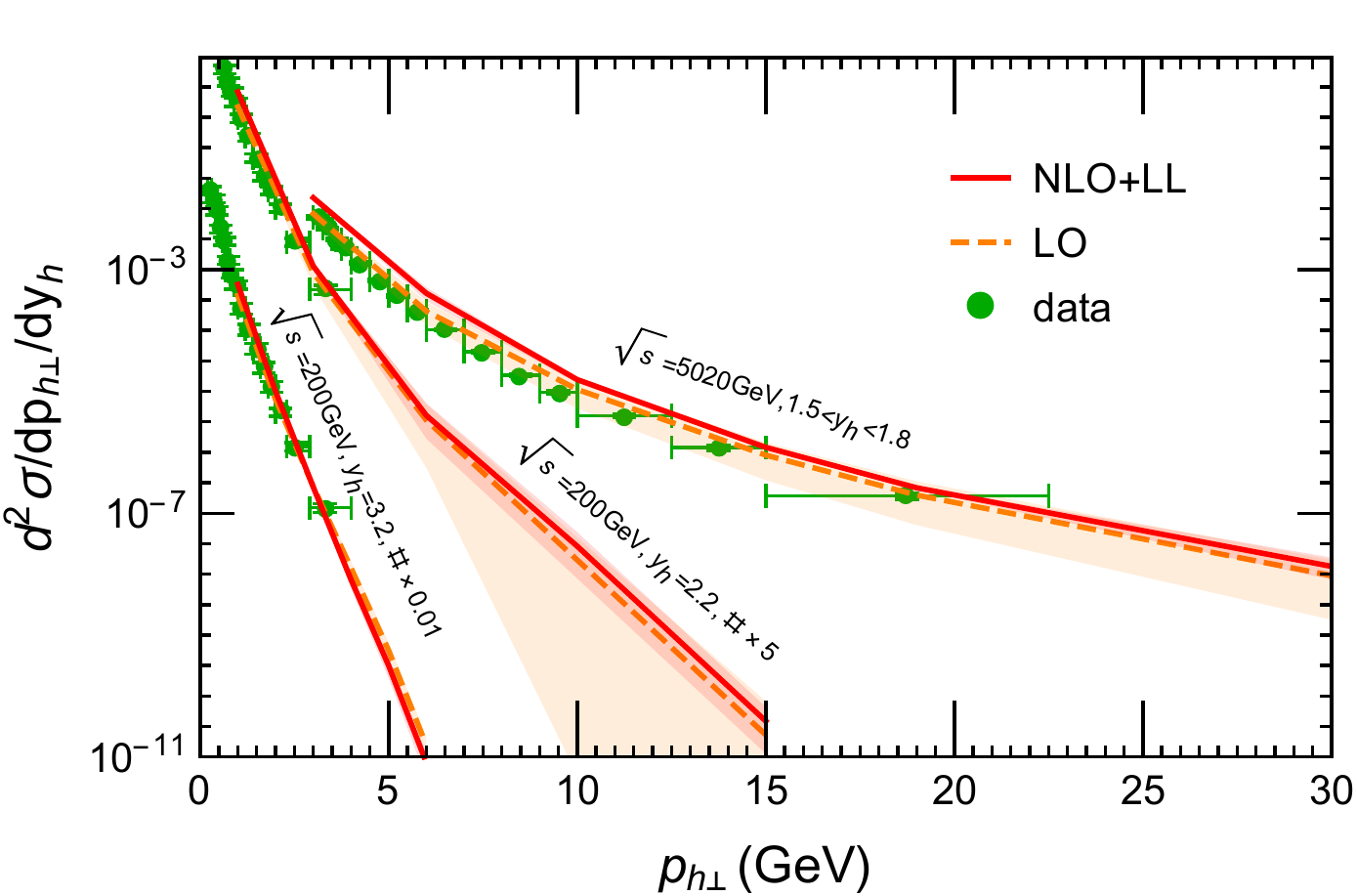}
\caption{Comparison of the CGC threshold resummation prediction with the LHC~\cite{ATLAS:2016xpn} and RHIC~\cite{BRAHMS:2004xry} data. Figure taken from Ref.~\cite{Liu:2020mpy}~\label{fig:small-x}}
\end{center}
\end{figure}

It is shown in~\cite{Liu:2020mpy} that  the formalism
\begin{itemize}
    \item reproduces the NLO results in covariant gauge. The soft function reproduces the so-called kinematic constraint, which was put in by hand in the previous calculations, but shows up naturally and automatically within the SCET/CGC hybrid formalism.
    \item systematically regulates all the divergences  by dimensional regularization or by the rapidity regulator. And the final results are consistent with the $\overline{\text{MS}}$ scheme; No cutoffs that violate the SCET power counting are introduced;
    \item naturally assigns the IR/rapidity poles to either the PDF/FFs or the nucleus CGC dipole distribution. The scale dependence correctly reproduces the DGLAP evolution for the proton and the final state hadron, as well as the small-$x$ BK equation for the colliding nucleus.
    \item allows for the resummation of the threshold logarithms, which formally can be achieved through the evolution factor of the jet and the soft function 
     \bea\label{eq:threshevo}
 {\bm U}_{{J}} & {\bm U}_{{S}}
=   \exp\left[  \, - 
 \frac{\alpha_s }{ \pi }  \int  \frac{\mathrm{d}x_\perp }{\pi}
 \left(  \ln \frac{\nu_S}{\nu_J} I_{BK,r}   + \ln \frac{X_f}{X_A}   
I_{BK}
\right) 
   {\bf T}^a_i  {\bf T}^{a'}_j  \, W_{aa'}(x_\perp ) 
 \right]   \,.
 \eea 
Here we see clearly that the resummation is not a conventional Sudakov evolution but it contains a complicated color structures and is highly non-linear in color. The exponentiation of the CGC Wilson line $W_{aa'}$ suggests that multiple structures will arise at higher orders which, again, cannot be reproduced by conventional threshold resummation. After the threshold resummation, the problem of the negative cross section is resolved and the theoretical predictions are found to agree with the known data~\cite{Liu:2020mpy}. A different approach to the negative cross section problem can be found in~\cite{Shi:2021hwx}.
    
\end{itemize}

The success of handling the threshold logarithms in forward $pA$ collisions suggests plausible future applications to other small-$x$ phenomena by combining the SCET and CGC formalism. For instance to apply the formalism to the forward di-jet/di-hadron production at the EIC. At the moment, the hybrid formalism is performed in a brute force way by coupling the CGC shock wave (Glauber Wilson line) to the SCET soft and collinear fields. In the future, a first principle and more systematic treatment is required along this line.

\section{Fragmentation and small-$\zh$ resummation}
\label{sec:fragmentation}

Another critical observable requiring resummation is fragmentation. Fragmentation is simply the inclusive cross-section for a particle to carry a certain amount of momentum in the event, while ignoring the rest of the process. These cross-sections have long been foundational to our understanding of both the perturbative and non-perturbative structure of high energy scattering. Perturbative physics since, at asymptotically high energies, the scaling properties with the hard momentum scale of the collision can be calculated, but non-perturbative physics plays a critical role since the observable is defined directly on the hadrons exiting the hard interaction. 

The canonical example of fragmentation is $e^+e^-\rightarrow h+X$, where one asks how many hadrons in the event carry a fraction of the total energy of the event. This example also enjoys one of the oldest factorization theorems of QCD, as well as the corresponding processes $ep\rightarrow h+X$ and $pp\rightarrow h+X$, see Refs.~\cite{Altarelli:1979kv,PhysRevD.18.3705,Collins:1981uw,Furmanski:1981cw}. If $Q$ is the center of mass energy of the electron-positron pair, and $\ph$ is the momentum of the hadron, the cross-section is: 
\begin{align}
    \zh&=\frac{2Q\cdot p_h}{Q^2}\,, \text{ with } 0\leq \zh \leq 1,\\
\label{eq:frag_fact}    \frac{1}{\sigma_0}\frac{d\sigma}{d\zh}&=\sum_{i=q,\bar{q},g}\int_{\zh}^{1}\frac{dz}{z}C_{i}\Big(z,\mu^2,Q^2\Big)d_{h/i}\Big(\frac{\zh}{z},\mu^2,\Lambda_{QCD}^2\Big)+\mathcal{O}\Big(\frac{\Lambda^2_{QCD}}{Q^2}\Big)\,.
\end{align}
Here $C_i$ is the coefficient function that describes how the $e^+e^-$ pair excite the QCD vacuum at high energies, and from this exits a parton of flavor $i$ carrying energy fraction $z$. From this parton, a hadron is formed carrying energy fraction $\zh$, and this process is captured by the fragmentation function $d_{h/i}$. The fragmentation function depends upon the scale of confinement $\Lambda_{QCD}$, and both functions depend on the factorization scale $\mu$. The cross-section or number density $d\sigma/\sigma_0$, (normalized by the total inclusive cross-section) does not depend on the factorization scale $\mu$, leading to the renormalization group equations:
\begin{align}
\bar{C}_{i}(n,\mu^2,Q^2)&=\int_{0}^{1}\frac{dz}{z}z^{n+1}C_{i}(z,\mu^2,Q^2)\,,\\
\bar{d}_{h/i}(n,\mu^2,Q^2)&=\int_{0}^{1}\frac{dz}{z}z^{n+1}d_{h/i}(z,\mu^2,\Lambda_{QCD}^2)\,,\\
    \mu^2\frac{d}{d\mu^2}\bar{C}_{i}(n,\mu^2,Q^2)&=-\sum_{j}\gamma^T_{ji}(n)\bar{C}_{j}(n,\mu^2,Q^2)\,,\\
    \mu^2\frac{d}{d\mu^2}\bar{d}_{h/i}(n,\mu^2,Q^2)&=\sum_{j}\gamma^T_{ij}(n)\bar{d}_{h/j}(n,\mu^2,Q^2)\,.
\end{align}
 All processes have the same fragmentation function $d_{h/i}$, an early and testable prediction of factorization: the universality of the hadronization process when the cross-section is sufficiently inclusive. Moreover, the fragmentation to hadrons can be generalized to fragmentation to jets, where one replaces the intrinsic cut-off scale of confinement with the jet radius $R$, as was done in Refs.~\cite{Dasgupta:2016bnd,Kang:2016mcy,Dai:2016hzf}. At fixed order, $C_{i}$ has been calculated to order $\alpha_s^2$ in Ref.~\cite{Rijken:1996vr,Rijken:1996ns,Mitov:2006wy}, with partial results at the same order for the DIS fragmentation process Refs.~\cite{Abele:2021nyo,Anderle:2016kwa}, and $\gamma^T_{ij}$ has been calculated to order $\alpha_s^3$ Refs.~\cite{Mitov:2006ic,Moch:2007tx,Almasy:2011eq,Chen:2020uvt}.

The cross-section is largest in the $\zh\rightarrow 0$ limit. That is, the bulk of hadrons produced have low energy. This has its origins in the fact that QCD enjoys an enhanced infra-red region of phase-space, and the $\zh\rightarrow 0$ limit probes the soft singularities of QCD. The importance of this region was recognized early on, Refs.~\cite{Mueller:1981ex,Mueller:1982cq,Bassetto:1982ma}, and a resummation of the logarithms of $\zh$ in the anomalous dimension was accomplished at leading and next-to leading order, however, with a scheme dependence which did not interface well with the minimal subtraction schemes of dimensional regularization, making it difficult to compare to fixed order calculations. At leading log order in the $\zh\rightarrow 0$ limit, which in mellin space corresponds to $n\rightarrow 0$, we have:
\begin{align}
   \gamma^T_{gg}(n)&=\frac{n}{4}\Big(-1+\sqrt{1+8\frac{\alpha_s C_A}{\pi n^2}}\Big)+...\,. 
\end{align}
After performing the inverse mellin transform, one can see then that this is double logarithmic series, \emph{in the anomalous dimension itself}.

The resummation was further developed to embody the concepts of angular-ordering and hard cut-offs in transverse momentum, culminating in the so-called mixed leading log (MLL) approximation, see Refs.~\cite{Fong:1990nt,Dokshitzer:1991ej,Khoze:1996dn}. The MLL has seen surprising phenomological success (Ref.~\cite{Akrawy:1990ha}) when coupled to the hypothesis of parton-hadron duality (\emph{very} crudely, each parton becomes a hadron), see Ref.~\cite{Azimov:1984np}, but with no clear connection to traditional fixed order calculations of the anomalous dimension or the coefficient function. 

More recently, two other approaches to small-$\zh$ resummation have been developed, which have been pushed beyond leading logarithmic order, and include a clear relationship to fixed order calculations.\footnote{Though see also Ref.~\cite{Albino:2011si}, which clarified the role of scheme-dependence in small-$\zh$ resummations, thus enabling the first successful matching of fixed-order to resummed perturbation theory in the context of fragmentation.} The first approach of Refs.~\cite{Kom:2012hd,Vogt:2011jv}, which has also been used to resum transverse-momentum dependent fragmentation functions Ref.~\cite{Luo:2020epw}, makes use of a set of recursion relations based on an ansatz that the energy fraction $\zh$ must ultimately cut-off soft divergences in the calculation. This leads to an over-determined set of linear relations, whose unique solution enables N$^2$LL resummation of the coefficient function, and N$^3$LL resummation of the anomalous dimension with arbitrary flavor dependence. Parton-to-pion fragmentation functions have been fit using these resummed coefficient functions and anomalous dimensions Ref.~\cite{Anderle:2016czy}. 

The second approach of Ref.~\cite{Neill:2020bwv} seeks to put the notion of an angular-ordered cascade on firmer theoretical grounds, by explicitly writing an evolution equation for the partonic coefficient function. The evolution equation seeks to make use of a universal kernel that in principle should be calculated from the BFKL equation that governs forward scattering and the resummation of the DIS cross-section as Bjorken $x$ goes to zero. Consistency of the BFKL equation \cite{Kuraev:1976ge,Lipatov:1976zz,Balitsky:1978ic,Kuraev:1977fs,Lipatov:1985uk} with the collinear factorization of DIS cross-section has been known to give the resummation of space-like anomalous dimensions of twist-two operators, where the consistency in dimensional regularization was first worked out in Ref.~\cite{Catani:1993ww,Catani:1994sq}. However, work on the resummation of the soft physics of NGLs in Refs.~\cite{Weigert:2003mm,Hatta:2008st,Avsar:2009yb,Caron-Huot:2015bja,Caron-Huot:2016tzz} established an essential connection between forward scattering physics and the resummation of non-global-logs (NGLs), based on the correspondence between the BK and BMS equations~\cite{Balitsky:1995ub, Kovchegov:1999yj,Banfi:2002hw}. While the NGL appear in semi-exclusion processes where radiation is vetoed, in certain angular regions Ref.~\cite{Neill:2020bwv} makes the connection between forward scattering and soft parton cascades explicit for the case of inclusive fragmentation, writing a deformation of the BFKL equation which resums the \emph{time-like} anomalous dimensions in complete analogy to the space-like case. It is worth noting that Refs.~\cite{Marchesini:2003nh,Marchesini:2004ne} also examined a form of the BFKL equation that could be used for fragmentation, but never explicitly connected the results with the traditional collinear factorization approach.

In addition to developing a BFKL approach to soft fragmentation, the authors also showed how one may write DGLAP-style evolution equations that resum the small-$\zh$ logs. Focusing on the case of pure Yang-Mills, one postulates that in $4-2\epsilon$ dimensions, the cross-section $d(\zh,\mu^2,Q^2,R^2)$ for fragmenting a gluon of energy fraction $\zh$ can be defined with an angular cut-off: the descendant parton cannot approach within angle $R$ of their parent or siblings. This regulates the collinear divergence. Then one takes the limit $R\rightarrow 0$ using the evolution equation:
{\small\begin{align}\label{eq:generalized_resummation_equation_for_DGLAP_momentum_space}
  R^2\frac{\partial}{\partial R^2} \zh^{1+2\epsilon} d\Big(\zh,\mu^2,Q^2,R^2\Big)&=\rho\Big(\frac{\mu^2}{R^2Q^2}\Big)\zh^{1+2\epsilon} d\Big(\zh,\mu^2,Q^2,R^2\Big)\nonumber\\
  &\qquad+\int_{\zh}^{1}\frac{dz}{z}P\Big(\frac{\zh}{z};\frac{\mu^2}{z^2R^2Q^2}\Big)z^{1+2\epsilon} d\Big(z,\mu^2,Q^2,R^2\Big)\,,\\
  P\Big(\frac{\zh}{z};\frac{\mu^2}{z^2R^2Q^2}\Big)&=\sum_{\ell=1}^{\infty} P^{(\ell-1)}\Big(\frac{\zh}{z};\alpha_s;\epsilon\Big)\,\Big(\frac{\mu^2}{z^2R^2Q^2}\Big)^{\ell\epsilon}\,,\label{eq:expansion_of_kernel_P_momentum_space}\\
  \rho\Big(\frac{\mu^2}{R^2Q^2}\Big)&=\sum_{\ell=1}^{\infty}\rho^{(\ell-1)}(\alpha_s;\epsilon)\Big(\frac{\mu^2}{R^2Q^2}\Big)^{\ell\epsilon}\,.
\end{align}}
$P$ is derived from the space-like anomalous dimension in $4-2\epsilon$ dimensions of Refs.~\cite{Ball:2005mj,Ciafaloni:2005cg,Ciafaloni:2006yk}, and $\rho$ is connected to the beta-function. After taking $R\rightarrow 0$, we recover the bare perturbative cross-section in $4-2\epsilon$ dimensions, so:
\begin{align}\label{eq:Factor_IR_Divergences_remind}
  \bar d\Big(n,\mu^2,\mu^2,0\Big)&=\text{exp}\Big(\int_{0}^{\alpha_s}\frac{d\alpha}{\beta(\alpha,\epsilon)}\gamma^T(\alpha,n)\Big)\,\mathcal{R}^T(\alpha_s,n).
\end{align}
Both $\gamma^T$ and $\mathcal{R}^T$ are resummed automatically in the soft limit, and indeed, $\mathcal{R}^T$ is the soft resummed coefficient function of Eq. \eqref{eq:frag_fact}. The evolution equation \eqref{eq:generalized_resummation_equation_for_DGLAP_momentum_space} can be viewed as an appropriate generalization of the reciprocity relations of Refs.~\cite{Dokshitzer:2005bf,Basso:2006nk}

 \begin{figure}\center
   \hspace{-10pt} \includegraphics[width=0.45\textwidth]{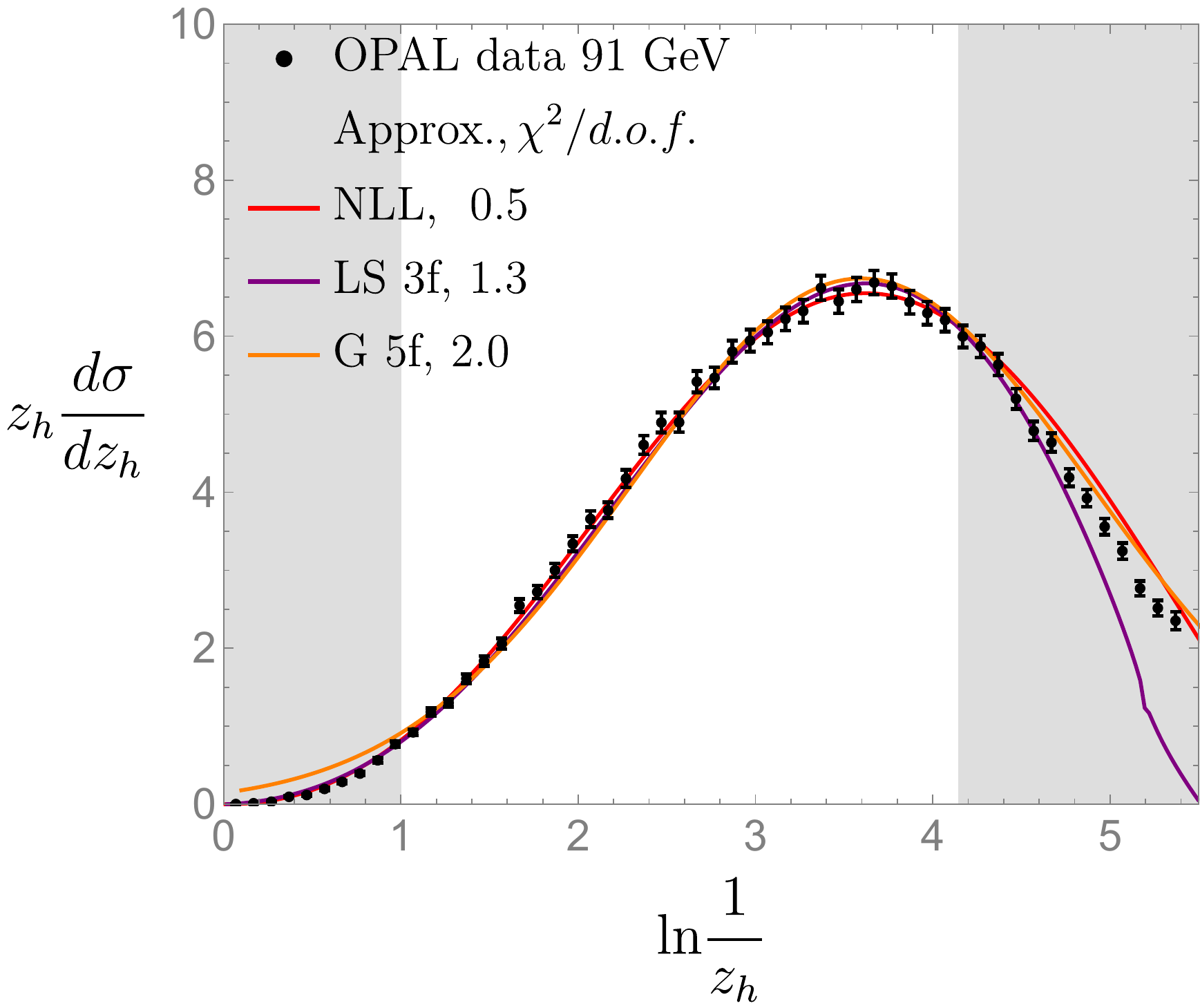}    \includegraphics[width=0.45\textwidth]{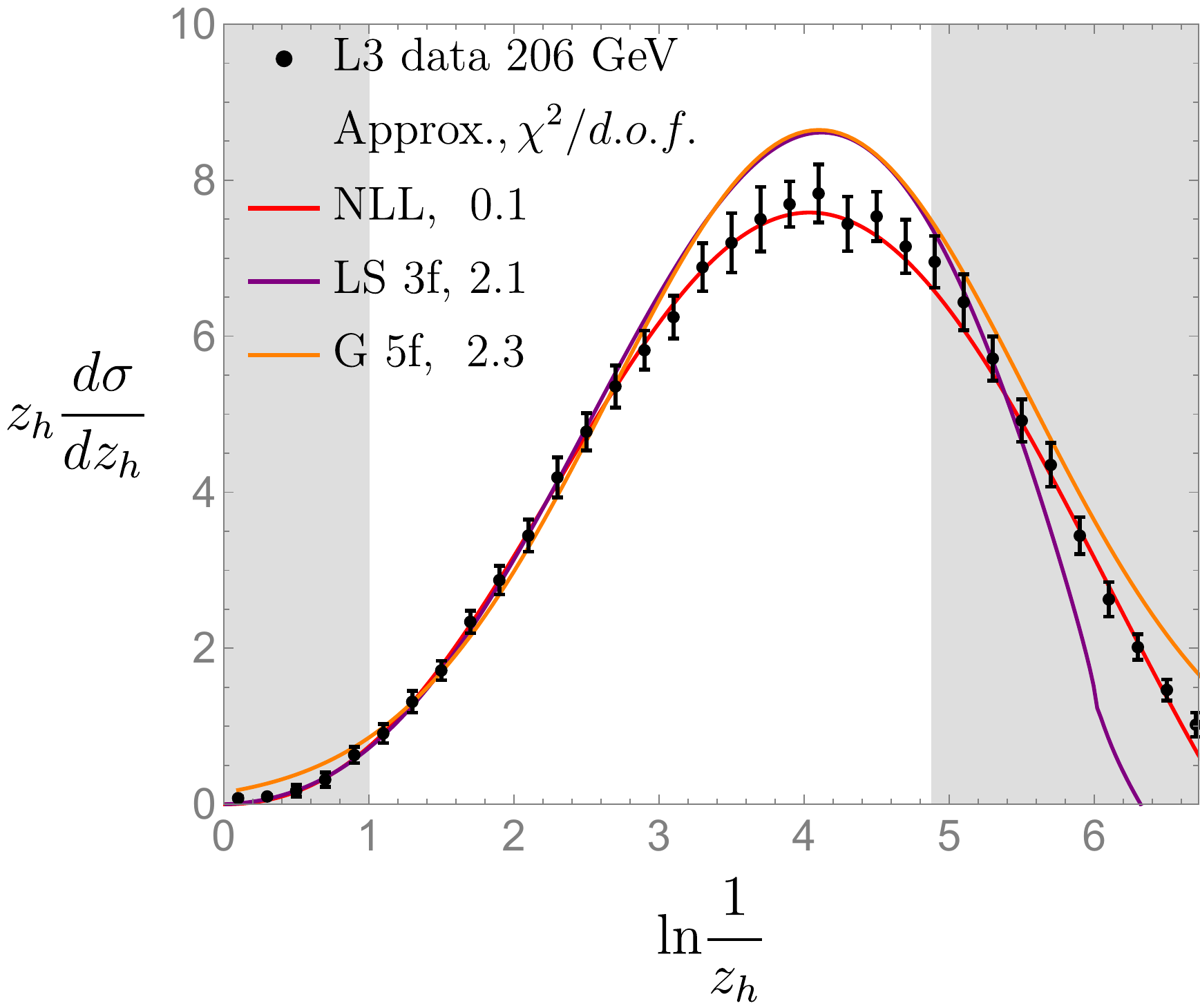}
   \caption{\label{fig:opal} The fragmentation spectrum for $e^+e^-\rightarrow h^{\pm}+X$ at center of mass energies $Q=91$ and $206$ GeV. The gray region is excluded from the $\chi^2$/d.o.f determination based on hadron mass corrections or $\zh\sim O(1)$, and the quoted $\chi^2$/d.o.f. is the goodness-of-fit for that data set. All curves are fitted at $91$ GeV and evolved to $206$ GeV. Compared to the NLL curve (red) is the evolved 3 flavor scheme MLLA limiting spectrum (purple LS) from Ref.~\cite{Akrawy:1990ha}, and a 5 flavor scheme evolving the moments of a Gaussian according to the MLLA anomalous dimension, as in Ref.~\cite{Fong:1990nt} (orange G). }
 \end{figure}  

Going into the future, there are multiple directions to study. Firstly, it is necessary to extend the resummation of soft fragmentation beyond the well-studied case of $e^+e^-\rightarrow h+X$ to processes that include initial state hadrons like $ep\rightarrow h+X$ and $pp\rightarrow h+X$. Secondly, there was a history of so-called fractal observables that were resummed in the unsystematic MLL approximation, many of which are outlined in Ref.~\cite{Khoze:1996dn}. It would be important to revisit these observables with an improved understanding of the resummation of the soft fragmentation region. Additionally, fragmentation in a constrained jet (Ref.~\cite{Procura:2009vm,Jain:2011iu}) is critical for many experimental analyses, as is following the flow of charge within the jet Ref.~\cite{Waalewijn:2012sv,Chang:2013rca,Li:2021zcf}. The extension of the soft resummation to these cases would bring insights into the flow of charge and momentum in QCD cascades. Lastly, the approach of Ref.~\cite{Neill:2020bwv} should be extended to include the full flavor structure of QCD, and the exact connection to forward scattering physics should move beyond conjectures and verification order-by-order, but involve full-scale proofs and examples showing the limits of the correspondence between time-like and space-like branching processes. 

 \begin{figure}\center
   \hspace{-10pt}  \includegraphics[width=0.45\textwidth]{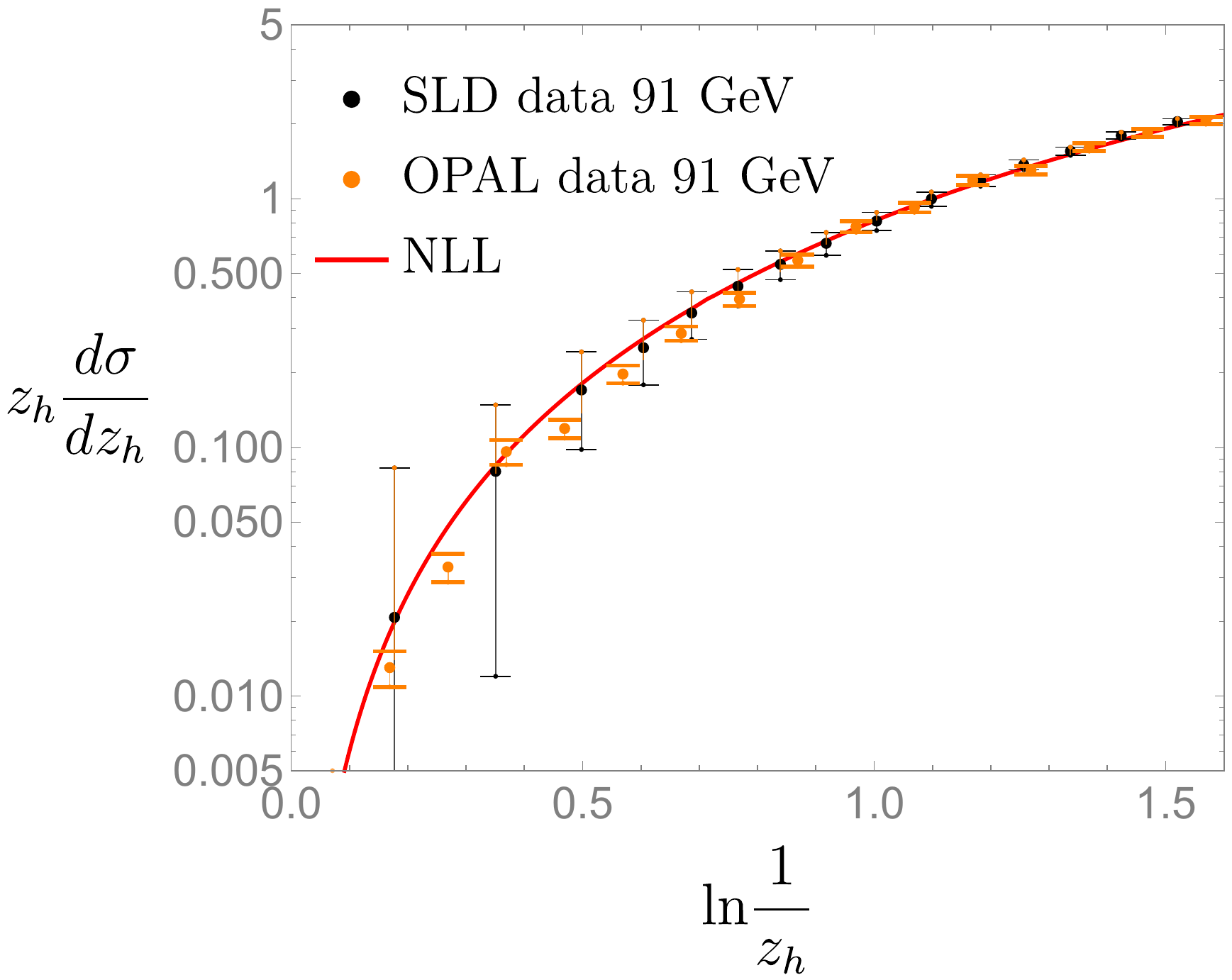}   \qquad \includegraphics[width=0.45\textwidth]{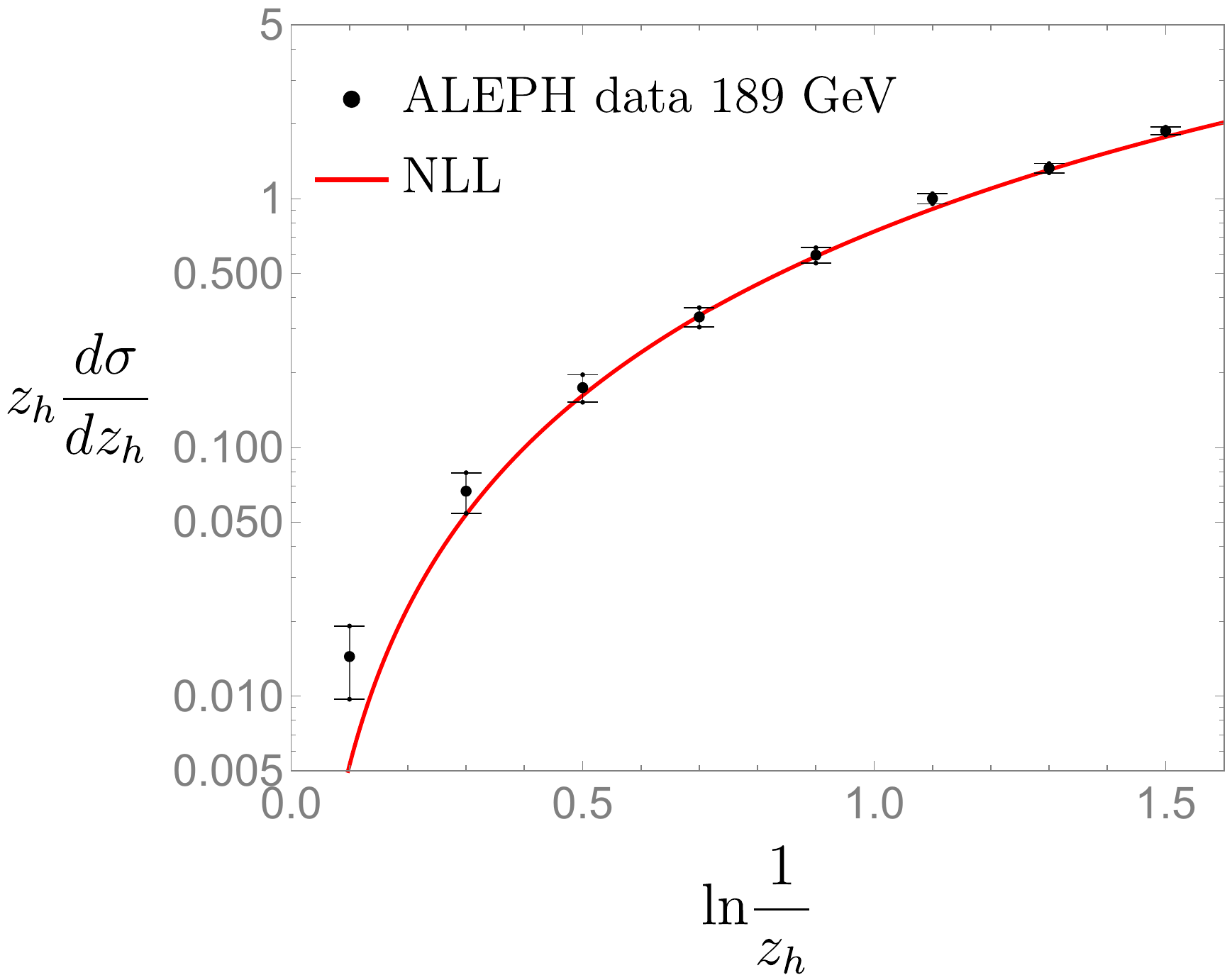}
   \caption{\label{fig:largezh} The fragmentation spectrum extrapolated into the large $\zh$ region for $e^+e^-\rightarrow h^{\pm}+X$ at center of mass energies $Q=91$ and $189$ GeV. The region below ln$\frac{1}{\zh}<$0.5 was excluded from the fit. Data taken from the OPAL, SLD, and ALEPH collaborations Refs.~\cite{Abe:2003iy,Akrawy:1990ha,Heister:2003aj}. The NLL result was a global fit to data from $Q=35$ to $206$ GeV.}
 \end{figure}  

On a more phenomological note, Ref.~\cite{Neill:2020tzl} initiated a study of parton-hadron duality based off of the improved understanding of the resummation of the soft physics from Ref.~\cite{Neill:2020bwv}, resumming all logarithms of $\zh$ to NLL order. The results showed a marked improvement on the older MLL approximation, see Fig.~\ref{fig:opal}, where fragmentation to charged hadrons is compared to theory predictions making use of parton-hadron-duality hypothesis. Perhaps even more surprising was the ability to describe the data in regions outside the soft limit, see Fig.~\ref{fig:largezh}. With improved analytical control of the perturbative structure of the theory, the time seems ripe to revisit parton-hadron duality, and the ultimate limits of perturbation theory.

\section{Conclusions}  \label{sec:conc}

The perturbative approach to calculate QCD scattering processes from first principles can be spoiled by the presence of large logarithmic corrections that appear in the power expansion of cross sections in the QCD strong coupling constant $\alpha_s$. These large logarithmic corrections need to be resummed to all orders in $\alpha_s$, which restores the predictive power of perturbative QCD. In this work, we reviewed recent progress in QCD resummation techniques. An improved understanding of the all-order structure of QCD is relevant for the high-luminosity era at the LHC as well as future experiments such as the Electron-Ion Collider. First, we considered the resummation at subleading power, which has become possible for several processes. For example, the threshold resummation for the Drell-Yan process and Higgs production via gluon fusion has been now been performed at subleading power and leading logarithmic accuracy. It was found that the numerical impact at the LHC can be comparable to high-precision resummation at leading power. We discussed techniques both in soft collinear effective theory and direct QCD. Both methods still face conceptual challenges which hinder systematic evaluation of the higher logarithmic orders. 
Second, we reviewed progress in the calculation of jet and jet substructure observables. Here the challenging aspect is the simultaneous resummation of multiple classes of large logarithmic corrections. Especially, jet substructure observables can be sensitive to several disparate energy scales that require careful treatment to identify the relevant energy scales and establish the corresponding factorization theorems. We discussed jet measurements at the LHC and outlined how jet physics can provide new insights into the structure of hadrons at the future EIC. Third, we focused on small-$x$ resummation, which is relevant in the high-energy limit where the momentum fraction carried by the initial-state gluon becomes small. We discussed recent work that combines resummation techniques in SCET and the Color Glass Condensate formalism. The additional resummation of threshold logarithms allows for the description of forward $pA$ scattering data. Lastly, we focused on the QCD fragmentation process. The inclusive hadron cross section peaks in the limit where the momentum fraction of the identified hadrons becomes small $z_h\to 0$. Large logarithmic corrections in the coefficient function and the time-like anomalous dimension need to be taken into account to achieve a reliable prediction within perturbative QCD. We discussed recent progress that employs space-time reciprocity relations, which connect the description of final-state hadrons to BFKL dynamics. The resummation of small-$z_h$ logarithms allows for a precise comparison to the available data from LEP and SLD colliders. The topics discussed here provide a starting point for future studies at high perturbative accuracy both at leading and subleading power. QCD resummation is a critical ingredient to stress test the Standard Model of particle physics, better understanding QCD itself, and to improve searches for physics beyond the Standard Model.

\acknowledgments
MvB acknowledges the UK Science and Technology Facilities Council (STFC) grant 452 number ST/T000864/1 and the Royal Society Research Professorship (RP$\setminus$R1$\setminus$180112). SJ is supported by the UK Science and Technology Facilities Council (STFC) under grant ST/T001011/1. XL is supported by the National Natural Science Foundation of China under Grant No.~12175016. DN  is supported by the U.S. Department of Energy under Contract No. 89233218CNA000001 and by the LDRD program at LANL. FR is supported by the Simons Foundation under the Simons Bridge program for Postdoctoral Fellowships at SCGP and YITP, award number 815892 and the NSF, award number 1915093. RS is supported by the United States Department of Energy under Grant Contract DE-SC0012704. LV is supported by Fellini, Fellowship for Innovation at INFN, funded by the European Union's Horizon 2020 research programme under the Marie Sk\l{}odowska-Curie Cofund Action, grant agreement no. 754496. GV is supported by the United States Department of Energy, Contract DE-AC02-76SF00515. JW is supported by the National Natural Science Foundation of China under Grant No. 12005117. 

\bibliographystyle{JHEP}
\bibliography{master}

\end{document}